\documentclass[%
 reprint,
 superscriptaddress,
 nofootinbib,
 amsmath,amssymb,
 aps,
 prx,
]{revtex4-2}

\usepackage[hidelinks]{hyperref}
\usepackage{graphicx}
\usepackage{dcolumn}
\usepackage{bm}
\usepackage{booktabs}
\usepackage{amsmath,amssymb,amsbsy,amstext, amsthm, amsfonts}
\usepackage{color}
\usepackage{slashed}
\usepackage[dvipsnames]{xcolor}
\usepackage{dsfont}
\usepackage{verbatim}
\usepackage[utf8]{inputenc} 
\usepackage{ragged2e}
\usepackage{mathtools}
\usepackage{xfrac}
\usepackage{blkarray}

\usepackage{tikz}
\usetikzlibrary{shapes.misc}
\usepackage[compat=1.1.0]{tikz-feynman}

\newcommand{\half}{\frac{1}{2}}

\def\Vhrulefill{\leavevmode\leaders\hrule height 0.7ex depth \dimexpr0.4pt-0.7ex\hfill\kern0pt}

\begin{document}

\title{Neutrino Masses from Generalized Symmetry Breaking}

\author{Clay C\'{o}rdova}
\affiliation{Enrico Fermi Institute, University of Chicago}
\affiliation{Kadanoff Center for Theoretical Physics, University of Chicago}
\author{Sungwoo Hong}
\affiliation{Enrico Fermi Institute, University of Chicago}
\affiliation{Theory Division, Argonne National Lab}
\affiliation{Department of Physics, Korea Advanced Institute of Science and Technology}
\author{Seth Koren}
\affiliation{Enrico Fermi Institute, University of Chicago}
\author{Kantaro Ohmori}
\affiliation{Faculty of Science, University of Tokyo}

\date{\today}

\begin{abstract}

\noindent
We explore generalized global symmetries in theories of physics beyond the Standard Model. Theories of $Z'$ bosons generically contain `non-invertible' chiral symmetries, whose presence indicates a natural paradigm to break this symmetry by an exponentially small amount in an ultraviolet completion.
For example, in models of gauged lepton family difference such as the phenomenologically well-motivated $U(1)_{L_\mu - L_\tau}$, there is a non-invertible lepton number symmetry which protects neutrino masses. 
We embed these theories in gauged non-Abelian horizontal lepton symmetries, e.g. $U(1)_{L_\mu - L_\tau} \subset SU(3)_H$, where the generalized symmetries are broken nonperturbatively by the existence of lepton family magnetic monopoles.
In such theories, either Majorana or Dirac neutrino masses may be generated through quantum gauge theory effects from the charged lepton Yukawas e.g. $y_\nu \sim y_\tau \exp(-S_{\rm inst})$.
These theories require no bevy of new fields nor ad hoc additional global symmetries, but are instead simple, natural, and predictive: the discovery of a lepton family $Z'$ at low energies will reveal the scale at which $L_\mu - L_\tau$ emerges from a larger gauge symmetry.
\end{abstract}

\maketitle

\makeatletter
\def\l@subsection#1#2{}
\def\l@subsubsection#1#2{}
\makeatother
\tableofcontents 


\section{Introduction}

With fifty years of the Standard Model behind us, it is clear we must look in as many directions as possible for new physics---from building ever-more-inventive experimental probes to understanding how subtle effects in quantum field theory might be connected to the real world.

Perhaps the surest sign that there is something else experimentally accessible is that we have already discovered physics beyond the Standard Model which interacts with the visible sector! Neutrino masses are not present in the Standard Model, and new dynamics \textit{must exist} to provide them. This new physics could be light, weakly coupled partners permitting Dirac masses, or it could be any variety of interactions inducing an effective dimension-five Majorana masses.  Regardless, new degrees of freedom are certainly out there to be discovered.

In this paper we will explore these issues from the viewpoint of novel generalized global symmetries \cite{Gaiotto:2014kfa}.
In particular, a prominent role will be played by so-called non-invertible symmetries which have recently been extensively investigated \cite{Tachikawa:2017gyf,Koide:2021zxj,Choi:2021kmx,Kaidi:2021xfk,Anosova:2022yqx,Roumpedakis:2022aik,Bhardwaj:2022yxj,Arias-Tamargo:2022nlf,Hayashi:2022fkw,Choi:2022zal,Kaidi:2022uux,Choi:2022jqy,Cordova:2022ieu,Antinucci:2022eat,Bashmakov:2022jtl,Damia:2022bcd,Moradi:2022lqp, Cordova:2022rer, Choi:2022rfe, Bhardwaj:2022lsg, Bartsch:2022mpm, GarciaEtxebarria:2022vzq, Apruzzi:2022rei, Delcamp:2022sjf, Heckman:2022muc, Freed:2022qnc, Niro:2022ctq, Kaidi:2022cpf, Antinucci:2022vyk,Bashmakov:2022uek} (See \cite{Cordova:2022ruw} for more complete review of these developments). We will use this perspective to build models of neutrino masses which are minimally generated by instantons and hence naturally small.  In particular, this constitutes a first example of generalized global symmetries having important implications for realistic theories of physics beyond the Standard Model. 

\subsection{Neutrino Masses}

Building on the successful discovery of neutrino oscillations by Super-Kamiokande \cite{Super-Kamiokande:1998kpq} and the Sudbury Neutrino Observatory \cite{SNO:2002tuh} around the turn of the millennium, the parameters of the neutrino sector are the subject of intensive experimental efforts.  We here briefly recall their coarse properties.

Precision measurements of the width of the $Z$ boson \cite{UA2:1987qqc,ALEPH:1989kcj} and the damping of large multipole anisotropies in the Cosmic Microwave Background \cite{Dunkley:2010ge,Keisler:2011aw,WMAP:2012nax} unambiguously dictate the existence of three `active' (by convention `left-handed') neutrinos. Observations of oscillations of flavor eigenstates on different length scales have pinned down two of the differences of mass eigenvalues as 
\begin{equation}
    m_2^2 - m_1^2  \simeq (9 \ {\rm meV})^2~, \quad |m_3^2 - m_i^2|  \simeq (50 \ {\rm meV})^2~,
\end{equation}
where the index $i=1,2$ above indicates that we do not know the full ordering, with either a `normal' $m_1 < m_2< m_{3}$ ($i=2$) or `inverted' $m_3 < m_1 < m_2$ ($i=1$) hierarchy allowed. 

The overall scale of neutrino masses has not been measured, but cosmological observations impose an upper bound on the sum of neutrino masses \cite{Lesgourgues:2006nd,Allison:2015qca,Palanque-Delabrouille:2015pga,Planck:2018vyg}, from the modification to the universe's evolution when cosmic background neutrinos become non-relativistic
:
\begin{equation}
\sum m_\nu \lesssim 150 \ \rm{meV}~.
\end{equation}
The prospect of measuring the mass scale with CMB Stage 4 (or alternatively by precise measurements of beta decay kinematics \cite{KATRIN:2021uub,KamLAND-Zen:2022tow}) this next decade remains enticing.  But for now, the data remain consistent with scenarios ranging from one exactly massless neutrino to all three neutrino masses at the same rough scale. 

Precise measurements of appearances and disappearances have also revealed that while the $m_1, m_2, m_3$ mass eigenstates are primarily composed of the electron, muon, tau neutrino respectively, their mixing is far larger than in the quark sector. In particular, the second and third generations are near `maximally mixed' ($\theta_{23}\sim \pi/4$).

\subsection{Hierarchies and Symmetries}

Even before asking the theoretical origin of the mixing structure, there is a question posed merely by the overall neutrino mass scale being parametrically below that of the electromagnetically-charged Standard Model fermions.  If neutrino masses have the same origin as the rest of the fermions, i.e.\ Yukawa couplings  to the Higgs field which generates mass by electroweak symmetry breaking, then this requires the introduction of minuscule coupling constants:
\begin{equation}
    \frac{y_\nu}{y_{\tau}}\sim 10^{-11}~,
\end{equation}
comparing the largest Yukawa among neutral leptons to that in the charged lepton sector. If neutrinos instead have Majorana masses through the irrelevant `Weinberg operator', this directly requires new physics which violates an exact global symmetry of the Standard Model at a large scale $\Lambda \sim 10^{14} \ \rm{GeV}$.

The problem of addressing hierarchies in effective field theory is closely related to an analysis of global symmetries \cite{tHooft:1979rat,Koren:2020biu}.  
Indeed, one way for a small parameter to manifestly be stable to radiative corrections is if, when the parameter exactly vanishes, a symmetry is restored.  
This feature is known as technical naturalness.  
In the case of neutrino masses, as with the rest of the Yukawa structure, technical naturalness ensures that Yukawa couplings which are small at some large scale will remain small upon evolving to the infrared.  
Then a fully satisfying explanation for their size may be postponed until high energies for our academic descendants to discover. 
Thus, in a sense, the problem of the overall neutrino mass scale is ultimately reduced to explaining the ultraviolet origin of small parameters. 
In broad strokes, our goal in the following is to evince that \textit{non-invertible} such protective symmetries motivate particularly minimal symmetry breaking mechanisms which generate exponentially small neutrino masses.

 \subsection{This Work}
 
We begin our analysis in Section \ref{sec:symbreaking} with a review of recent developments in generalized symmetry with a focus on non-invertible chiral symmetries \cite{Choi:2022jqy,Cordova:2022ieu}.  In particular, we discuss natural symmetry breaking mechanisms described in \cite{Cordova:2022ieu} arising from loops of monopoles.

 We then apply these ideas to phenomenological models working up in scale starting from the Standard Model in the infrared.  While the Standard Model itself does not enjoy any non-invertible symmetries,\footnote{In the approximation where fermion masses are neglected there are such non-invertible symmetries which can be used to understand aspects of pion physics \cite{Choi:2022jqy}.} we will show that such generalized global symmetries may play a role in understanding the phenomenology of well-motivated theories beyond the Standard Model. In particular, a careful analysis of the Standard Model symmetries in Section \ref{sec:SMsyms} leads us to consider theories of gauged lepton family differences $U(1)_{L_i - L_j}$, and in Section \ref{sec:intermediate} we show that gauging this Standard Model global symmetry results in non-invertible symmetries which protect neutrino masses. This suggests that such models have a natural ultraviolet completion in which small neutrino masses arise from the breaking of a one-form global symmetry by dynamical lepton family monopoles.  

Indeed, we exhibit ultraviolet completions of these theories where either Majorana or Dirac neutrino masses arise originally from instantons of a gauged non-Abelian `horizontal' lepton symmetry.\footnote{The idea of generating neutrino masses via instanton effects has been previously explored in string theory models using D-brane instantons \cite{Ibanez:2006da,Blumenhagen:2006xt,Antusch:2007jd,Cvetic:2008hi,Hoshiya:2021nux,Kikuchi:2022bkn}.  There, the ultraviolet physics responsible for mass generation is controlled by the size of the cycle wrapped by the brane, where one can probe extra dimensions.  It would be interesting to connect our general paradigm to these models perhaps by interpreting them in terms of non-invertible symmetries involving the Ramond-Ramond gauge fields.} These models are remarkably predictive: for example in our Dirac model below, the instantonic origin of neutrino masses can be discovered at the IR scale of the $Z'$ boson. In this case the UV scale $v_\Phi$ at which instantons are generated is determined by the neutrino mass scale and the measurements of $M_{Z'}$ and the coupling strength $\alpha_{\mu\tau}(M_{Z'}^2)$, 
\begin{equation}
    v_{\Phi}^2 \sim M_{Z'}^2 \left(\frac{m_\nu}{m_\tau}\right)^{\sfrac{3}{2}} \exp{\frac{3 \pi}{4  \alpha_{\mu\tau}(M^2_{Z'})}}~. 
\end{equation}

In the Dirac case we will find that, after adding right-handed neutrinos, their Yukawa interactions with the Higgs are protected by a non-invertible symmetry. In Section \ref{sec:dirModel}, we give an embedding of this theory in a gauged horizontal lepton symmetry $SU(3)_H$ which breaks the non-invertible symmetry only in the quantum theory via instantons, providing Dirac masses which are exponentially suppressed compared to those for charged leptons.
In the Majorana theory with solely the Standard Model light fermion content, there is a non-invertible lepton number symmetry whose breaking may arise from an embedding of $U(1)_{L_\mu - L_\tau}$ in a gauged $SU(2)_H \times U(1)_Z$, as we will show in Section \ref{sec:majModel}. 

\begin{figure}[h]
    \centering
    \begin{tikzpicture}
    \begin{feynman}
        \draw[fermion] (0,0) --node[midway,above=.1] {$\nu$} ++(1.5,0) coordinate(b);
        \draw[double,Green,thick] (b) arc (180:120:.5) coordinate(b1) arc(120:60:.5) coordinate(b2) arc (60:-180:.5);
        \draw[fermion] (b) ++ (1,0) coordinate(c)  ++(1.5,0) -- node[midway,above=.1] {$\nu$} (c);
        \draw[fill] (c) circle(.07);
        \draw[fill] (b) circle(.07);
        \node[anchor=north] at (2,-.6) {monopole};
        \draw[scalar] (b1) -- ++ (120:.5) node[cross out,draw,rotate=30,inner sep=0,minimum size=5] (x);
        \draw[scalar] (b2) -- ++ (60:.5) node[cross out,draw,rotate=60,inner sep=0,minimum size=5] (y);
        \node[anchor=east] at (x) {$\langle H\rangle$};
        \node[anchor=west] at (y) {$\langle H\rangle$};
    \end{feynman}
    \end{tikzpicture}
    \caption{An intuitive illustration of Majorana neutrino masses generated by monopole/dyon loops in our models. From an ultraviolet perspective the sum over monopole loops is identified with an instanton process.}
    \label{fig:mon_loop}
\end{figure}
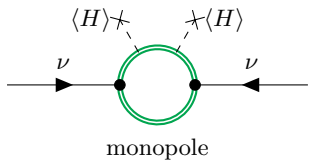

Thus in both these theories, the hierarchically low scale of neutrino masses is explained by a global symmetry which is classically respected but broken quantum mechanically. Rather than requiring complicated model-building to explain why either explicit or spontaneous breaking of \textit{ad hoc} symmetries in the neutrino sector are surprisingly small, here the ultraviolet gauge theory automatically provides a small breaking through the instanton action.
These effects can also be intuitively understood as loops of monopole states which correct the neutrino propagator (See Figure \ref{fig:mon_loop}).  This correspondence between monopole/dyon loops and quantum corrections due to instantons was decribed quantitatively in the context of axions in \cite{Fan:2021ntg}.  
In summary, in our models neutrinos become massive because they interact with the Higgs boson through the exchange of virtual monopoles.

Hence, the presence of a non-invertible symmetry in the theory portends the existence of this mechanism for neutrino masses satisfying Dirac's criteria for naturalness: order one numbers in the ultraviolet Lagrangian.  

Our work suggests a number of threads for future investigation. While we focus on the natural origin of the hierarchically small overall neutrino mass scale, we will make brief comments below about generating also the full mixing matrix while flowing down from the UV scale. However, a full exploration of integrating this mechanism into the large body of results on neutrino sector `textures' and related $L_\mu - L_\tau$ phenomenology is one interesting direction for future work.  

A further direction is to consider the physics of the lepton family monopoles, which to our knowledge have received quite little attention in the literature \cite{Bibilashvili:1990qm,Chkareuli:1991tp,Dvali:2021uvk}.  In the early universe, these particles may arise as interesting dynamical objects which may have unitarity-limited inelastic interactions with Standard Model leptons, yet which safely decay away after further symmetry-breaking. Such monopoles at the large gauge couplings that our models require may also see appreciable production at a lepton collider, and it would be prudent to understand their signatures.  

\section{Generalized Symmetry}\label{sec:symbreaking}

In this section we review recent developments regarding symmetry in quantum field theory.  We focus on the particular topic of non-invertible chiral symmetry \cite{Choi:2022jqy,Cordova:2022ieu} relevant for our phenomenological applications.  

\subsection{Symmetry Defect Operators }\label{sec:defectops}

The conceptual basis of recent progress is a broad interpretation of Noether's theorem which links topological operators to symmetries \cite{Gaiotto:2014kfa}.  To set the stage, recall that 
Noether's theorem connects a continuous global symmetry with a current operator $J^\mu$ with vanishing divergence:
\begin{equation}
    \partial_\mu J^\mu = 0~.
    \label{conservation}
\end{equation}
The Noether charge at time $t$ is constructed as an integral over all space
\begin{equation}
    Q(t) = \int_{\mathbb{R}^3} J^0(t,x)\mathrm{d}^3x~, \quad \frac{\mathrm{d}}{\mathrm{d}t}Q(t) = 0~,
\end{equation}
where conservation of the charge follows from \eqref{conservation}.

The idea of a topological operator is to consider a more general three-dimensional surface $\Sigma$ in spacetime (rather than just $\mathbb{R}^3$). This generalizes the Noether charge to an extended operator $Q$ with support on $\Sigma$:
\begin{equation}
    Q[\Sigma] \equiv \int_\Sigma J^\mu \mathrm{d}^3S_\mu~,
    \label{eqn:charge}
\end{equation}
where above $\mathrm{d}^3S_\mu$ is the vector volume element pointing in the normal direction to $\Sigma$.\footnote{In expressions such as \eqref{eqn:charge}, one superficially encounters a contraction of indices, however in fact the operator $Q[\Sigma]$ does not depend on the metric. This can be clearly seen for instance by reexpressing the defining integrals using differential forms. }

What does the local conservation law \eqref{conservation} tells us about the charge $Q[\Sigma]$? When we have a smooth deformation $\Sigma'$ of $\Sigma$,
the difference of the charges measured on $\Sigma'$ and $\Sigma$ can be computed by Stokes' theorem:
\begin{equation}
    Q[\Sigma'] - Q[\Sigma] = \int_{V} \partial_\mu J^\mu \mathrm{d}v~,
\end{equation}
where $V$ is the four-volume between $\Sigma'$ and $\Sigma$: $\partial V = \Sigma' - \Sigma$, and $\mathrm{d}v$ is the scalar volume element. Thus, the local conservation law \eqref{conservation} tells us that $Q[\Sigma]$ is invariant under smooth deformations of $\Sigma$.   In this case we say that the extended operator $Q[\Sigma]$ is \emph{topological}.

\begin{figure}[h]
    \centering
        \begin{tikzpicture}[baseline=-2]
            \draw (0,0)  node[circle,
                       shade,
                       ball color=orange,
                       minimum size=1.5cm,opacity=.5]{};
            \draw[fill] circle (.05);
            \node[anchor=south west] at (0,0) {$\mathcal{O}$};
            \node[anchor = south east] at (-.5,.5) {$Q_\alpha[\Sigma]$};
        \end{tikzpicture}
         $\quad=q\, \mathcal{O}$
    \caption{The charge operator $Q[\Sigma]$ wraps a local operator $\mathcal{O}(x)$ in spacetime.  Using the topological property, $Q[\Sigma]$ can be shrunk and results in a factor of the charge $q$ of $\mathcal{O}(x)$. }
    \label{fig:zeroform}
\end{figure}
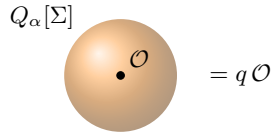

The smooth changes of the manifold $\Sigma$ do not extend to deformations which cause $\Sigma$ to cross local operators.  In this case the correlation functions of $Q[\Sigma]$ in general change.  More specifically,  when the three-surface $\Sigma$ surrounds a local operator $\mathcal{O}$ in spacetime, it measures the charge carried by the local operator. (See Figure \ref{fig:zeroform}.)
This follows from the Ward-Takahashi identity:
\begin{equation}
    \langle \partial_\mu J^\mu(x) \mathcal{O}(y) \cdots \rangle = q \delta^4(x-y)\, \langle \mathcal{O}(y) \cdots \rangle~,
\end{equation}
where $q$ is the charge of $\mathcal{O}$ and $\cdots$ represents other insertions away from $x$.

The charge operator $Q[\Sigma]$ can be viewed as generating the infintesimal (algebra) action of the symmetry.  The symmetry defect operator $U_\varphi[\Sigma]$ is simply the exponentiation of $Q[\Sigma]$ and hence corresponds to a finite (group) action of the symmetry:
\begin{equation}
    U_\varphi[\Sigma] = e^{i\varphi Q[\Sigma]}~.
\end{equation}
Note that such operators are now also labelled by a phase $e^{i\varphi}$ in $U(1)$ indicating the angle of charge rotation.  

One advantage of symmetry defect operators (as opposed to charges) is that the symmetry defect operators also exist for discrete symmetries where there is no corresponding conserved current.  In general for each element $g\in G$ of the symmetry group $g$ we have corresponding operator $U_g[\Sigma]$:
\begin{equation}
    g \in G \rightsquigarrow U_g[\Sigma]~.
\end{equation}
Intuitively, one may understand the extended operator $U_g[\Sigma]$ as prescribing discontinous field configurations along $\Sigma$, where the field values on the two sides of $\Sigma$ differ by the action of $g$. Alternatively, one may consider the defects in a phase where the associated symmetry is spontaneously broken.  In that case $U_g[\Sigma]$ is a domain wall connecting distinct vacua. The topological property of $U_g[\Sigma]$ is a manifestation of conservation of the symmetry, and generalizes Noether's theorem to the discrete setting. 

\subsection{One-Form Symmetry and Monopoles}\label{oneformmag}

We have seen that Noether's theorem can be recast as the existence of topological operators corresponding to symmetry operations. Adopting this point of view broadly then suggests that any operator which has topological correlation functions (except when crossing other operators) can be viewed as defining a kind of symmetry. In general for instance, a topological operator may have support which is lower-dimensional and gives rise to a so-called higher-form global symmetry \cite{Gaiotto:2014kfa}. 

\begin{figure}[h]
    \centering
    \begin{tikzpicture}[scale = 0.7,baseline = 0]
        \draw[->] (-1.2,-.5) -- ++(0,1) node[anchor = south] {$t$};
        \draw[->,Green] (0,-1.5) -- (0,0);
        \draw (0,0) ellipse (.6 and .2);
        \draw[Green] (0,0) -- (0,1.5);
        \node[anchor = south west] at (.5,.25) {$U_\varphi^M$};
        \node at (-.4,1.5) {$T$};
    \end{tikzpicture}
    $=e^{i\varphi}$
    \begin{tikzpicture}[scale = 0.7,baseline = 0]
        \draw[->,Green] (0,-1.5) -- (0,0);
        \draw[Green] (0,0) -- (0,1.5);
        \node at (-.4,1.5) {$T$};
    \end{tikzpicture}
    \caption{The action of magnetic one-form symmetry operator $U_\varphi^M$ on a 't Hooft line $T$ (the worldline of a heavy probe monopole). The symmetry operator is spatially placed so that it wraps the heavy monopole in a time slice. Since $U_\varphi^M$ is topological, it can be shrunk towards the monopole worldline, giving the phase $e^\mathrm{i\varphi}$. This is the magnetic version of Gauss' law and generalizes the action in Figure \ref{fig:zeroform}. }
    \label{fig:mag_oneform}
\end{figure}
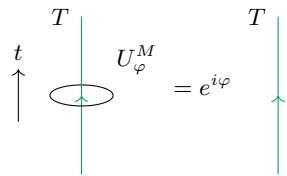

In this a key role will be played by one-form global symmetry which naturally arises in gauge theory.  These are symmetries where the charges are topological surface operators that act on line defects (physically, the worldlines of heavy charged particles). The most prominent example occurs in the familiar context of free Maxwell theory, i.e.\ $U(1)$ gauge theory with gauge field $A$ without charged matter where the equation of motion and the Bianchi identity can be recast as conservation of two distinct currents each with two vector indices:
\begin{equation}\label{oneformdiv}
    \partial^{\mu}J^E_{\mu\nu}=0~, \hspace{.2in} \partial^{\mu}J^M_{\mu\nu}=0~,
\end{equation}
where 
\begin{equation}
    J^E_{\mu\nu} = \frac1{4\pi e^2}F_{\mu\nu}~, \hspace{.2in} 
    J^M_{\mu\nu} = \frac1{32\pi^2}\varepsilon_{\mu\nu\rho \sigma}F^{\rho \sigma}.
    \label{eq:one-form_current}
\end{equation}
As is the case of ordinary symmetry, we can define the corresponding symmetry defect operator by
\begin{equation}
    U^E_{\varphi}[\Sigma_2] = e^{i\varphi\int_{\Sigma_2} J^E_{\mu\nu} \mathrm{d}S^{\mu\nu}}~, \quad 
    U^M_{\varphi}[\Sigma_2] = e^{i \varphi\int_{\Sigma_2} J^M_{\mu\nu} \mathrm{d}S^{\mu\nu}}~,
\end{equation}
where now these operators are supported on the two-dimensional manifold $\Sigma_2$, and electromagnetic action is $S_\text{EM}= \frac{1}{4e^2}\int F_{\mu\nu}F^{\mu\nu}\mathrm{d}^4x$.
The surface element tensor $\mathrm{d}S^{\mu\nu}$ is defined by $n_1^\mu n_2^\nu \mathrm{d}S$ where $n_1,n_2$ are the orthogonal normal vectors on $\Sigma_2$, and $\mathrm{d}S$ is the scalar surface element.
For example, when the surface $\Sigma_2$ is contained in a time slice $t=t_0$, we can take $n_1^{\mu} = \delta^{\mu 0}
$, and we have
\begin{equation}
    U^E_{\varphi}[\Sigma_2] = e^{i\frac{\varphi}{4\pi e^2}\int_{\Sigma_2} \mathbf{E}\cdot\mathrm{d}\mathbf{S}}~, \quad 
    U^M_{\varphi}[\Sigma_2] = e^{i \frac{\varphi}{16\pi^2}\int_{\Sigma_2} \mathbf{B}\cdot\mathrm{d}\mathbf{S}~},
\end{equation}
in the vector analysis notation.
These electric and magnetic one-form symmetry operators act on Wilson or 't Hooft lines by linking. See Figure \ref{fig:mag_oneform}.

The fact that the charged objects for one-form symmetries are worldlines of infinitely massive source particles also underlies the natural symmetry breaking method for these currents.  By introducing dynamical i.e.\ finite mass electric (magnetic) charges, sources can be screened by vacuum polarization and the underlying symmetries are broken.  At the more mechanical level, the divergences of the one-form symmetry currents \eqref{oneformdiv} are then non-zero operators.  

The simple physical picture of screening described above belies a subtle conclusion.  In effective field theory one cannot break higher-form global symmetries without changing the degrees of freedom.  For example, modifying the free Maxwell Lagrangian by higher-derivative terms:
\begin{equation}
    \mathcal{L}\supset \frac{1}{\Lambda^{4}}(F_{\mu\nu}F^{\mu\nu})^{2}+\cdots~,
\end{equation}
merely deforms the electric one-form symmetry current: 
\begin{equation}
    F_{\mu\nu}\rightarrow F_{\mu\nu}+\frac{1}{\Lambda^{4}}F_{\mu\nu}(F)^{2}+\cdots~,
\end{equation}
but leaves the rank of the symmetry unchanged since the right-hand side is conserved as a consequence of the equation of motion. 
This should be contrasted with standard symmetry breaking of ordinary (zero-form) symmetries where a charged local operator in the action violates conservation.  Instead, in the higher-form case, to any finite derivative order in effective field theory, the symmetry is unbroken. That is, there are no local operators charged under this symmetry.

For the magnetic one-form global symmetry, this conclusion is particularly dramatic: breaking this symmetry requires finite action field configurations carrying magnetic charge.  In the simplest case these are monopoles.  Within weakly-coupled effective field theory with coupling $g$ one may estimate the size of such symmetry breaking effects by modeling the monopole as arising from a Higgsing process at a scale $v_{\Phi}$.  Then the mass of the monopole and cutoff $\Lambda$ are parametrically given by:
\begin{equation}
    m_{\text{mon}}\sim \frac{v_{\Phi}}{g}~, \hspace{.2in} \Lambda \sim g v_{\Phi}~.
\end{equation}
The quantum vacuum contains such monopoles which propagate for a short proper time $\delta t \sim 1/\Lambda$ and hence gives rise to terms in the effective action which scale exponentially 
\begin{equation}
    \delta \mathcal{L}\sim \exp\left(-S_\text{mon}\right)\sim\exp\left(-m_{\text{mon}} \delta t\right)\sim \exp\left(-\#/g^{2}\right)~,
    \label{eq:mon_inst}
\end{equation}
where $S_\text{mon}$ is the one-particle action of the monopole and we have used the Schwinger representation of the monopole propagator.
Note that the scaling with coupling is the same as that of instanton corrections from a non-abelian group.  Indeed, the first quantized picture of loops of monopoles can often be traded for a sum over instanton sectors in an ultraviolet non-abelian group \cite{Fan:2021ntg}.

In summary, violations of magnetic one-form symmetry are naturally related to exponentially suppressed instanton-like corrections to the effective action.  For instance, this analysis applies to the $U(1)_{Y}$ hypercharge gauge group in the Standard Model which has an associated magnetic one-form global symmetry. 

\subsection{Non-Invertible Chiral Symmetry}\label{sec:noninv}

Our discussion has linked exponentially small corrections in an effective action to ultraviolet violations of magnetic one-form symmetry.  However, the pattern of these corrections is so far unclear: are these tiny corrections merely extra contributions to process which are already present in the low energy effective field theory, or are they the leading terms governing some processes?

Non-invertible symmetry provides a key tool to understand this essential question.  For the purposes of this work, we will confine our attention to those non-invertible symmetries which semi-classically appear as ordinary symmetries which are violated only by abelian instanton configurations.  The paradigmatic example is a classical symmetry encoded by a current $J_{\mu}$ which is violated by an abelian Adler-Bell-Jackiw (ABJ) anomaly: 
\begin{equation}\label{ABJabelian}
    \partial^{\mu}J_{\mu}=\frac{k}{32\pi^{2}}F_{\alpha\beta}F_{\gamma \delta}\varepsilon^{\alpha\beta \gamma \delta}~, \hspace{.2in} k\in \mathbb{Z}~,
\end{equation}
where $F$ is an \emph{Abelian} gauge field strength, and $k$ is an integral anomaly coefficient.  There are several immediate pragmatic conclusions from this equation:
\begin{itemize}
    \item A $\mathbb{Z}_{k}$ subgroup of the symmetry generated by $J$ is unaffected by the anomaly and remains as a standard (invertible) symmetry.
    \item Despite the presence of the non-trivial anomaly, the symmetry generated by $J$ is still preserved at the level of local operator correlation functions (and hence S-matrix elements). This follows from the fact that, when the gauge group is Abelian, the instanton processes needed to generate a net violation of the charge do not exist in simple configurations with only local operator insertions. Their absence follows straightforwardly from $\pi_3\left(U(1)\right) = 0$, so that gauge transformations cannot non-trivially wrap the boundary of 4d Euclidean spacetime. 
\end{itemize}

In spite of these points, the Abelian anomaly \eqref{ABJabelian} is not innocuous.  The most significant question is how to construct symmetry defects, the dimension three operators which perform finite chiral symmetry transformations defined in section \ref{sec:defectops}.  Ordinarily one would merely integrate the current as in \eqref{eqn:charge}, but the non-zero divergence in \eqref{ABJabelian} obstructs this naive prescription.  

As recently discussed in \cite{Choi:2022jqy,Cordova:2022ieu}, the correct construction of symmetry defects for symmetries suffering from Abelian ABJ anomalies involves coupling the bulk degrees of freedom to a non-trivial three-dimensional topological field theory supported along the defect worldvolume $\Sigma$.  These degrees of freedom are anyonic particles with fractional spin and abelian statistics.  They couple to the bulk via their one-form magnetic symmetry which is gauged by $F$. 

In somewhat more detail, consider a finite symmetry rotation by angle $2\pi/kN$ where $N\in \mathbb{Z}$.  This is one $N$-th of the angle $2\pi/k$ whose corresponding symmetry defect is unaffected by the anomaly.  Then, the worldvolume degrees of freedom on the symmetry defect are an Abelian gauge field $C$ with Chern-Simons level $N$ and action:
\begin{equation}
   \mathcal{L}= \frac{iN}{4\pi}\int_{\Sigma} C_{\mu} \partial_{\nu}C_{\sigma}\varepsilon^{\mu\nu\sigma}d^{3}x +\frac{i}{2\pi}\int_{\Sigma} C_{\mu}\partial_{\nu}A_{\sigma}\varepsilon^{\mu\nu\sigma}d^{3}x~.
\end{equation}
More generally, the construction above may be carried out for any finite rotation by a rational angle.  Below we sometimes abuse notation and still refer to such a symmetry for all rational angles as a $U(1)$, since it enforces the same selection rules on correlation functions of local operators.

The fact that the defect now supports a non-trivial quantum field theory means, in modern terminology, that the symmetry with an Abelian ABJ anomaly has become non-invertible.  This has two closely related effects.  

\begin{figure}[h]
    \centering
        \begin{tikzpicture}[scale = .9,baseline = 10]
            \draw[Green,->] (0,0) arc (-90:270:1 and .7);
            \draw (0,.7)  node[circle,
                       shade,
                       ball color=orange,
                       minimum size=2.3cm,opacity=.5]{};
                       
            \node[anchor=north] (0,0) {$T$};
            \node[anchor=south east] at (-1,1.3) {$\mathcal{D}_{kN}$};
        \end{tikzpicture}
    $\quad = \quad$
    \begin{tikzpicture}[scale = .9, baseline = 10]
            \draw[Green,->,fill=blue,fill opacity=.4] (0,0) arc (-90:270:1 and .7);
            \node[anchor=north] at (0,0) {$TW^{1/kN}$};
            \node at (0,.7) {$U^M_{2\pi /kN}$};
    \end{tikzpicture}
    \caption{When the non-invertible symmetry defect $\mathcal{D}_{kN}$ for the chiral symmetry wraps a loop of 't Hooft line, it induces a fractional $\frac1{kN}$ electric charge on the 't Hooft line. Such a dyon with fractional charge has to be attached to the electromagnetic dual of a Dirac string whose world volume is identified with $U_{2\pi /kN}^M$. This indicates that dynamical monopoles break the chiral symmetry.}
    \label{fig:D_on_T}
\end{figure}
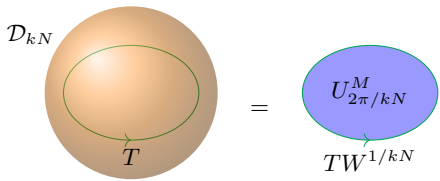

First, when acting on 't Hooft lines (worldlines of heavy magnetic monopoles), the symmetry defect acts to give them a fractional electric charge (via the Witten effect) and hence converts such line operators to open surface operators where the surface supports an integral of the magnetic one-form symmetry current $F_{\mu\nu}$. (See Figure \ref{fig:D_on_T}.) This is a consequence of the fact that magnetic charges can activate the divergence in \eqref{ABJabelian}.

Second, a finite rotation, followed by a rotation by an inverse angle does not result in the identity operator.  Instead, this composition leaves behind a ``condensate"  of one-form symmetry operators.  In equations, let $\mathcal{D}_{kN}(M)$ be symmetry defect associated to the angle $2\pi/kN$, where $M$ is the three-manifold (spatial slice) which supports the defect, and let $\overline{\mathcal{D}}$ be the operator generating a rotation by the opposite angle.  Upon colliding these operators we find the result:\footnote{In formulas such as \eqref{fusionalg} one often encounters topological invariants such as homology/cohomology which superficially vanish if the spacetime manifold or $M$ is taken to be sufficiently simple.  However, in general the insertion of other (extended) operators can be viewed as creating topology.  Thus, even in flat space one will see the sum over surface operators provided we consider a sufficiently general correlation function. }
  \begin{eqnarray}
        \mathcal{D}_{kN}(M)\times \overline{\mathcal{D}}_{kN}(M)&\sim & \sum_{\substack{{\text{two-cycles}} ~ S}} \exp\left(\frac{2\pi \mathrm{i}}{N}\int_{S}J^M_{\mu\nu}\mathrm{d}S^{\mu\nu}\right)~,  \nonumber \\
        &\sim&
        \sum_{\substack{{\text{two-cycles}} ~S}} U_{{2\pi}/{N}}^M[S]~,\label{fusionalg}
    \end{eqnarray}
    where $J^M_{\mu\nu}$ is the magnetic one-form symmetry current introduced in \eqref{eq:one-form_current}.
    The right-hand side is a sum over insertions of magnetic one form symmetry defects wrapping two-dimensional cycles $S\subset M$.\footnote{In 
\eqref{fusionalg} we ignore phases in the sum which depend on the triple self-intersection number of $S$, which play no role in this paper.}

    Hence the symmetry operator  $\mathcal{D}_{kN}(M)$ in general does not admit an inverse.  We will presently see the consequences of equation \eqref{fusionalg} in our models below.
    
\subsection{Chiral Symmetry Breaking by Monopoles}

The utility of the preceding discussion is that it provides a natural link between symmetries which enjoy abelian ABJ anomalies (and are hence non-invertible) and magnetic one-form symmetry breaking effects.  To illustrate this let us discuss possible mechanisms for breaking a non-invertible chiral symmetry in effective field theory.  We imagine that at long distances, the non-invertible symmetry is approximately respected by the physics, while at shorter distance scales there are symmetry violating effects.  We may then contemplate two broad possibilities:
\begin{itemize}
    \item One may directly add a chiral symmetry violating local operator $\mathcal{O}(x)$ to the Lagrangian:
    \begin{equation}
         L \supset \int d^{4}x~\frac{1}{\Lambda^{n}}\mathcal{O}(x)~.
    \end{equation}
    Then, as in the violation of any ordinary global symmetry, the current $J_{\mu}$ acquires a classical contribution to its divergence breaking the symmetry.
    \item One may violate the magnetic one-form symmetry.  In this case as remarked in subsection \eqref{oneformmag} one expects exponentially small corrections to the effective action scaling as instanon contributions $\exp(-\#/g^{2})$.  The algebra \eqref{fusionalg} links the non-invertible symmetry to the magnetic one-form symmetry and hence implies that these terms will be lead to exponentially small violations of the non-invertible chiral symmetry. 
\end{itemize}

It is the second of these mechanisms that is particular to a non-invertible symmetry.  At a pragmatic level it implies that a model with a non-invertible symmetry comes naturally equipped with a mechanism to violate the symmetry by exponentially small effects: the propagation of magnetic monopoles.  This is to be contrasted with standard symmetry violation by higher-dimensional charged operators whose size depends on the details of the ultraviolet.  Instead in the case where the non-invertible symmetry violation is communicated by violating the magnetic one-form symmetry the physics is universal.


\section{Symmetry of the Standard Model} \label{sec:SMsyms}

{\setlength{\tabcolsep}{0.6 em}
\renewcommand{\arraystretch}{1.3}
\begin{table}[h]\centering
\large
\begin{tabular}{|c|c|c|c|c|c|c|c|}  \hline
 & $  Q_i  $ & $ \bar u_i $ & $ \bar d_i $ & $  L_i  $ & $ \bar e_i $ & $  N_i  $ & $ H $ \\ \hline
$SU(3)_C$ & $\mathbf{3}$ & $\bar{\mathbf{3}}$ & $\bar{\mathbf{3}}$ & -- & -- & -- & --\\ \hline
$SU(2)_L$ & $\mathbf{2}$ & -- & -- & $\mathbf{2}$ & -- & -- & $\mathbf{2}$\\ \hline
$U(1)_{Y}$ & $+1$ & $-4$ & $+2$ & $-3$ & $+6$ & -- & $-3$\\ \hline
$U(1)_{B}$ & $+1$ & $-1$ & $-1$ & -- & -- & -- & --\\ \hline
$U(1)_{L}$ & -- & -- & -- & $+1$ & $-1$ & $-1$ & --\\ \hline 
\end{tabular}\caption{Representations of the Standard Model Weyl fermions under the classical gauge and global symmetries. We normalize each $U(1)$ so the least-charged particle has unit charge. We list also the charges of the right-handed neutrino $N$ and the Higgs boson $H$.  }\label{tab:charges}
\end{table}}

We now begin our construction of neutrino mass models protected by non-invertible symmetries.  We start in the infrared with the Standard Model and subsequently increase the energy scale.  Thus, our first task is to review the global symmetries of the Standard Model.  As emphasized above, we are particularly interested to understand classical symmetries which are broken by quantum effects \cite{Koren:2022bam,Wang:2022eag}. 
 
 The Standard Model is a gauge theory with gauge group:\footnote{There is a well-known ambiguity in the global structure of this gauge group and we may replace $G_{SM} \mapsto G_{SM}/\Gamma$, where $\Gamma$ is a subgroup of $\mathbb{Z}_6$. This ambiguity will not play any role in the discussion to follow. }
 \begin{equation}
     G_{SM} = SU(3)_C \times SU(2)_L \times U(1)_Y~,
 \end{equation}
 with the familiar Weyl fermion representations shown in Table \ref{tab:charges}. Here, the index $i=1, \cdots, N_{g}$ labels the number of generations (or families) of each type of matter field.  (In practice we will often leave $N_{g}$ as a variable in formulas to follow though nature has chosen $N_{g}=3$.)   A systematic way to understand the global symmetries is to first consider only the effect of the kinetic terms $\bar \Psi_i \slashed{D} \Psi_i$ for the matter fields and then sequentially take into account interactions and quantum effects.  
 
With only kinetic terms, we can rotate the families amongst each other 
\begin{equation}
    \Psi_i \longrightarrow U_{ij}\Psi_i ~,
\end{equation} 
 where $U_{ij}\in U(N_{g})$ is a unitary matrix.  This results in a $U(N_g)^5$ classical global symmetry of the fermion gauge covariant kinetic terms.

The Higgs field $H$ and associated mass-generating Yukawa couplings drastically reduce the symmetry of the Standard Model fermions. 
Given some assignment of charges under a global symmetry, we can always use our freedom to add $U(1)_Y$ charges to set the Higgs charge to zero.
We use this convention in the following.  The structure of these interactions is 
\begin{equation}\label{eqn:SMyuk}
    \mathcal{L} \supset y^u_{ij} \tilde{H} Q_i \bar u_j + y^d_{ij} H Q_i \bar d_j + y^e_{ij} H L_i \bar e_j~,
\end{equation}
where $\tilde{H} \equiv i \sigma_2 H^\star$.  The observed Yukawa matrices $y$ (equivalently, the fermion masses and flavor changing processes) explicitly break all of the non-Abelian continuous global symmetries, as they provide different masses for the generations.

To elucidate the classical symmetry preserved by the interactions \eqref{eqn:SMyuk}, we must take into account field redefinitions that we may use to simplify the couplings.  In the lepton sector, independent rotations of $L_i$ and $\bar e_i$ enable us to diagonalize the Yukawa matrix:
\begin{equation}
    y^e_{ij} H L_i \bar e_j \longrightarrow y_i^e H L_i \bar e_i~.
\end{equation}
So the lepton Yukawa interaction links together the transformations of the left- and right-handed leptons, but the so-called `lepton family symmetries' remain good classical symmetries.  We therefore have separate phase rotations for each generation:
\begin{equation}
    U(1)_{e}\times U(1)_{\mu} \times U(1)_{\tau}~,
\end{equation}
under which $L_i$ and $\bar e_i$ transform oppositely. (We sometimes also refer to these lepton family symmetries as $U(1)_{L_{i}}.$) In particular, the conventional lepton number $U(1)_{L}$ is the minimal linear combination of symmetries above which acts identically on each family.

In the quark sector, the Standard Model has right-handed partners for both the up and down quarks. This means we cannot simultaneously diagonalize both $y^u$ and $y^d$ while preserving $SU(2)_{L}$ .  Hence the would-be `quark family symmetries' are explicitly broken.  The only remaining global symmetry in the quark sector is therefore an overall $U(1)_B$ quark number called `baryon number.'  Note that with the charge assignments in Table \ref{tab:charges}, a $U(1)_{B}$ rotation by an $N_{c}$-th root of unity can be compensated for by a transformation in the center of the color gauge group:  
\begin{equation}
    \exp{\left(\frac{2\pi iB}{3}\right)}\in SU(3)_{C}~.
\end{equation}
Therefore the classical global symmetry of the quark sector is in fact $U(1)_{B}/\mathbb{Z}_{N_{c}}$.

\begin{figure}[h]
    \centering
        {\includegraphics[clip, trim=0.0cm 0.0cm 0.0cm 0.0cm, width=0.35\textwidth]{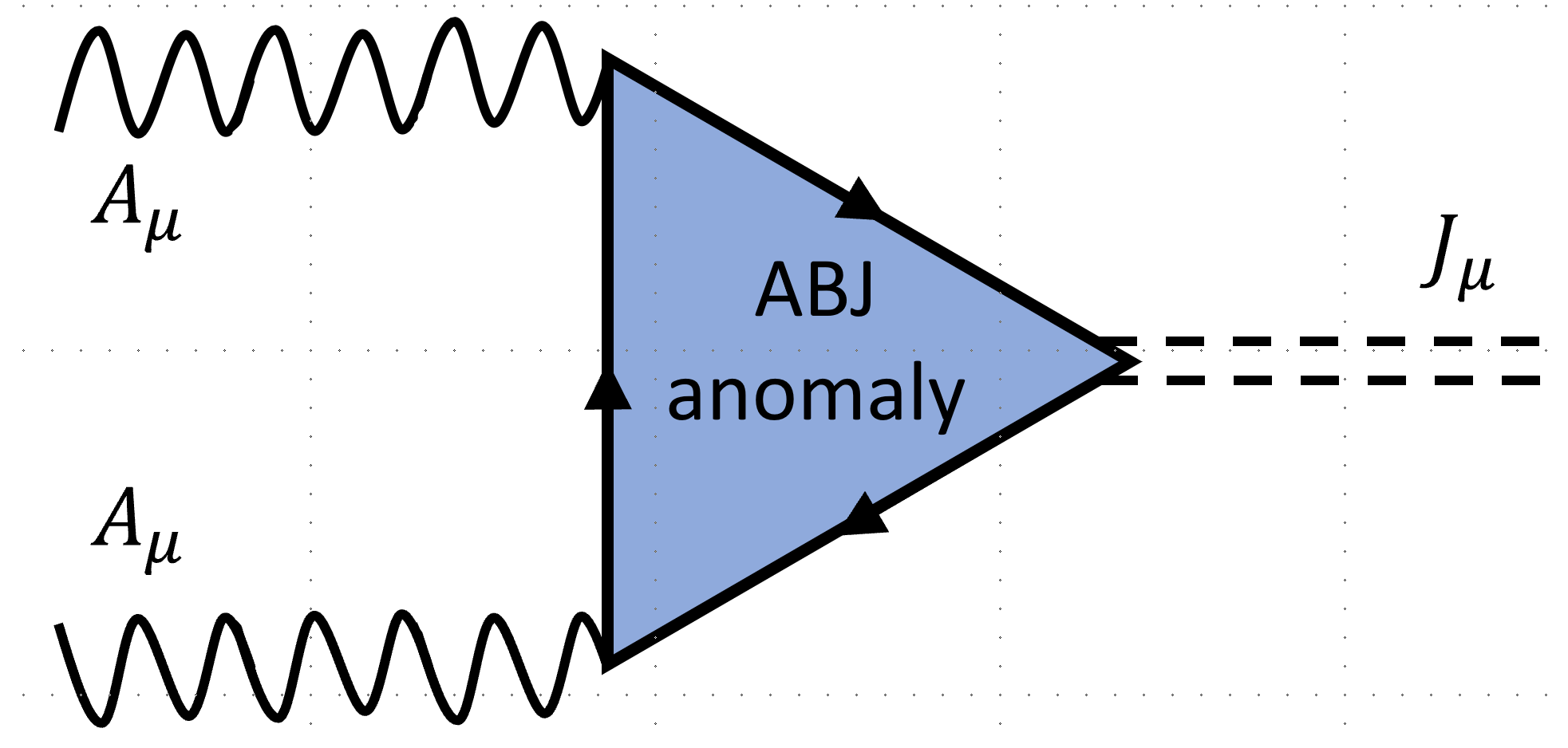}}
    \caption{A triangle diagram which breaks current conservation of the classical symmetries  of the Standard Model. }
    \label{fig:ABJDiag}
\end{figure}

In summary then, the Yukawas leave an abelian classical symmetry group:
\begin{equation}\label{classicalsym}
    \text{Classical Symmetry}\cong  U(1)_{e}\times U(1)_{\mu} \times U(1)_{\tau}\times \frac{U(1)_{B}}{\mathbb{Z}_{3}}~,
\end{equation}
and we must now investigate which of these classical symmetries survive quantum effects. The relevant triangle diagrams are illustrated schematically in Figure \ref{fig:ABJDiag}.
As we are interested in terms that break the classical symmetry, we focus here on those terms which give a non-trivial operator-valued divergence to the classically conserved currents: 
\begin{equation}
    \partial^{\mu}J_{\mu}=\frac{c_{L}}{32\pi^{2}}\mathrm{Tr}(W_{\mu\nu} W_{\rho \sigma})\varepsilon^{\mu\nu\rho \sigma}+\frac{c_{Y}}{32\pi^{2}}B_{\mu\nu} B_{\rho \sigma}\varepsilon^{\mu\nu\rho \sigma}~,
\end{equation}
where above $J_{\mu}$ indicates any of the currents of \eqref{classicalsym}, $W_{\mu\nu}$ is the field strength of the $SU(2)_{L}$ weak gauge group, and $B_{\mu\nu}$ is the field strength of the $U(1)_{Y}$ hypercharge gauge group.  Table \ref{tab:anomSM} summarizes the resulting anomaly coefficients.
\begin{table}[h]\centering
\large
\begin{tabular}{|c|c|c|c|c|}  \hline
 & $SU(2)_L^2$ & $U(1)_Y^2$ & $SU(3)_{c}^{2}$  \\ \hline

$U(1)_B$ & $N_g N_c$ & $-18 N_g N_c$ & $0$ \\ \hline 

$U(1)_{L_k}$ & $1$ & $-18$ & $0$   \\ \hline

$U(1)_L$ & $N_g$ & $-18 N_g$ & $0$  \\ \hline

\end{tabular}\caption{Anomaly coefficients of classical global symmetries in the Standard Model.  $L=L_{e}+L_{\mu}+L_{\tau}.$}\label{tab:anomSM}
\end{table}
Each of these anomaly terms leads to distinct consequences:
\begin{itemize}
    \item Non-zero $c_{L}$: The selection rules for $J_{\mu}$ are violated by instantons in the weak sector.  This means that the symmetry is broken from a $U(1)$ factor to a discrete group $\mathbb{Z}_{c_{L}}$ controlled by the anomaly coefficient. 
    \item Non-zero $c_{Y}$: The selection rules for $J_{\mu}$ are not violated since there are no Abelian instantons in trivial spacetime topology.  As discussed in section \ref{sec:noninv}, such an anomaly with $U(1)_Y$ alone leads to non-invertible chiral symmetries. 
\end{itemize}

The dynamical violation of selection rules in the case of non-zero $c_L$ is seen explicitly via the appearance of  't Hooft vertices in the effective action. These are multi-fermion operators generated by instantons that explicitly violate the anomalous symmetry.  The appearance of fermions in the 't Hooft vertex arises because $SU(2)_{L}$ charged fermions necessarily have zero modes in the presence of an $SU(2)_{L}$ instanton.  Grassmann statistics then force the instanton contribution to the correlator to vanish unless fermion fields are inserted to saturate all of the zero modes (see e.g.\ \cite{Morrissey:2005uza, Koren:2022ezi} for further pheno discussion).

The overall scale of these corrections is controlled by the action of the instantons $\exp(-S) \simeq \exp(- 8 \pi^2/g_{L}^2)$. Thus, for the electroweak sector they are only relevant in the hot early universe where the thermal instanton configurations go `on-shell' as sphalerons and are closely tied to electroweak baryogenesis.

For later applications we emphasize that 't Hooft vertices are a feature of non-abelian gauge dynamics and instantons.  By contrast for Abelian gauge fields no corresponding instanton process exists.

In our context a given $U(1)$ global symmetry will have mixed anomalies with both $SU(2)_{L}$ and $U(1)_{Y}$.  In this situation, non-invertible symmetry can arise only when there are transformations (continuous or discrete) that are unaffected by $SU(2)_{L}$ instantons, but which have non-vanishing $U(1)_{Y}^{2}$ anomalies.  It is straightforward to check that for the Standard Model matter content this scenario does not occur. Indeed, the anomaly coefficient of any global $U(1)$ with $U(1)_{Y}^{2}$ is always an integer multiple of that with $SU(2)_{L}^{2}$. This means that any global $U(1)$ rotations not violated by $SU(2)_L$ instanton effects are necessarily free of $U(1)_Y$ anomaly.  Thus, while $U(1)_Y$ does have a magnetic one-form symmetry, the Standard Model does not have any non-invertible symmetries---their would-be effects are always swamped by those of $SU(2)_L$.

Examining the anomaly coefficients in Table \ref{tab:anomSM} we can now deduce the true global symmetry of the Standard Model. The condition of vanishing $SU(2)_{L}^{2}$ anomaly implies a single linear relation on the charges, leaving three continuous $U(1)$ factors of \eqref{classicalsym} preserved.  These may be taken to be the difference of lepton family symmetries as well as a non-anomalous combination of lepton and baryon number.  In total then the Standard Model has symmetry:
\begin{equation}\label{truesym}
    U(1)_{L_{e}-L_{\mu}}\times  U(1)_{L_{\mu}-L_{\tau}}\times \frac{U(1)_{B-N_{c}L}}{\mathbb{Z}_{N_{c}}}~.
\end{equation}
With an eye towards later developments, we also note that there is a discrete $\mathbb{Z}_{N_{g}}^{L}$ subgroup of $U(1)_{L}$  which survives as a global symmetry.  This is contained in \eqref{truesym} since on all fields:
\begin{equation}\label{z3subs}
    \exp\left(\frac{2\pi i}{3}L\right)=\exp\left(\frac{2\pi i}{3}\left((L_{e}-L_{\mu})-(L_{\mu}-L_{\tau})\right)\right)~.
\end{equation}

The Standard model field content and global symmetries described in this section imply that neutrinos are exactly massless.  In particular, the symmetry \eqref{truesym} prevents Majorana neutrino masses from being generated by the effective dimension five Weinberg operator:
\begin{equation}
   \mathcal{L}\supset \frac{y^N_{ij}}{\Lambda} (\tilde{H} L_i)(\tilde{H} L_j)~,
\end{equation}
even though this operator is gauge invariant. Alternatively, if we add right-handed neutrinos $N_i$ then we may directly include Yukawa couplings to generate neutrino masses: 
\begin{equation}
    \mathcal{L}\supset y^N_{ij} \tilde{H} L_i N_j~.
\end{equation}
In this case, the masses respect the factors in \eqref{truesym} that involve overall lepton number, but for generic couplings $y_{ij}^{N}$ violate the family difference symmetries $L_{i}-L_{j}$ in the same manner as the quark Yukawas $y^u, y^d$. 

Thus, neutrino masses---whether Majorana or Dirac---imply that some portion of the global symmetry \eqref{truesym} is approximate and hence that our understanding of their ultraviolet fate is incomplete.

\section{Symmetry of Leptophilic $Z'$ Models} \label{sec:intermediate}

We now describe the implications of gauging an additional $U(1)$ factor, or in particle physics terminology the existence of a $Z'.$ We denote the gauge field by $A_{\mu}'$ with field strength $F_{\mu\nu}'.$ We will see that such models lead to non-invertible symmetry and provide natural mechanisms constraining neutrino physics.   

The simplest possible scenario explored here is to gauge a $U(1)$ subgroup of the symmetry \eqref{truesym} of the Standard Model.  In order for such a gauging to be consistent, we must still take care that the cubic 't Hooft anomaly vanishes for the new dynamical $U(1).$ With the strict Standard Model field content, this singles out the lepton family difference symmetries $L_{i}-L_{j}$ as those to potentially gauge.  With only slightly more complexity, we may also add right-handed neutrinos and then gauge $B-N_cL.$  

Below, we describe features of gauged lepton family models generally without reference to an ultraviolet embedding.  In later sections we will realize such models from non-abelian gauge theory.

\subsection{Gauged $L_{\mu}-L_{\tau}$ } \label{sec:gaugedMuTau}

Consider gauging a single combination of lepton family difference $L_{i}-L_{j}$.  Note that while mathematically any choice of $i$ and $j$ is allowed, gauging $L_\mu - L_\tau$ is both least constrained and most well-motivated. To the former point, effects of new interactions with first generation charged matter are far more easily probed and have a correspondingly larger set of constraints (see e.g.\ \cite{Heeck:2011wj,Wagner:2012ui,Elahi:2015vzh,Bauer:2018onh,Chun:2018ibr}). In particular, in the regime of large gauge coupling which will be picked out by our models, for a representative $g_{\mu\tau} \sim 1$, the $Z'$ mass must be above $M_{Z'} \gtrsim 1 \ \rm{TeV}$ \cite{Drees:2018hhs,Huang:2021nkl}. 
To the latter point, models with gauged $L_\mu - L_\tau$ have seen much study the past two decades as a potential explanation for experimental anomalies seen in precise measurements of $(g-2)_\mu$ (e.g. \cite{Ma:2001md,Baek:2001kca,Harigaya:2013twa,Biswas:2016yan,Lindner:2016bgg,Kamada:2018zxi}) and in $B$ meson branching ratios (e.g. \cite{Altmannshofer:2014cfa,Crivellin:2015mga,Altmannshofer:2016jzy}), as well as for the structure of neutrino mass matrices. In general, gauged lepton family symmetries are a well-motivated extension which tie together signatures in a wide variety of frontiers, from colliders \cite{Capdevilla:2021rwo,delAguila:2014soa,Liu:2018xsw,Ekhterachian:2021rkx}, to cosmology \cite{Escudero:2019gzq,Araki:2021xdk}, to direct detection \cite{Heeck:2010pg}, to astrophysics \cite{Araki:2014ona,KumarPoddar:2020kdz,Dror:2019uea}, to the intensity \cite{Kaneta:2016uyt,Shimomura:2020tmg,NA64:2022rme,Cesarotti:2022ttv}, and precision \cite{Asai:2021wzx,Ibe:2016dir} frontiers. 

\begin{table}[h]\centering
\large
\begin{tabular}{|c|c|c|c|c|}  \hline
 & $U(1)_{B}$ & $U(1)_{L_{e}}$  & $U(1)_{L_{\mu}}$&$U(1)_{L_{\tau}}$ \\ \hline

$U(1)_{L_{\mu}-L_{\tau}}^{2}$ & $0$ & $0$ & $1$ &1\\ \hline

\end{tabular}\caption{Mixed anomalies of classical global symmetries of the Standard Model  with a gauged $U(1)_{L_{\mu}-L_{\tau}}$.}\label{tab:anomSM2}
\end{table}

 When we promote $L_\mu - L_\tau$ to a gauge symmetry we must revisit the fate of the global symmetries of the Standard Model.  Here we consider models with no additional light fields beyond those of the Standard Model which contribute to the anomaly analysis.  
 
 Each of the classical global symmetries now has a new anomaly coefficient written in Table \ref{tab:anomSM2}.  Taking into account the $SU(2)_{L}^{2}$ anomalies leading to \eqref{truesym}, we find that one linear combination of currents is gauged, another is fully anomaly free and remains as a standard (invertible) global symmetry, while the final linear combination has trivial $SU(2)_{L}^{2}$ anomaly but non-trivial $U(1)_{L_{\mu} - L_{\tau}}^{2}$ anomaly coefficient: hence it becomes a non-invertible symmetry of this class of models.  We enumerate each of these linear combinations in Table \ref{tab:globMuTau}. 
{\setlength{\tabcolsep}{0.6 em}
\renewcommand{\arraystretch}{1.3}
\begin{table}[h]\centering
\large
\begin{tabular}{|c|c|}  \hline
Gauged & $U(1)_{L_{\mu}-L_{\tau}}$ \\ \hline

Invertible & $\sfrac{U(1)_{B-N_g N_c L_e}}{\mathbb{Z}_{N_c}}$ \\ \hline

Non-Invertible & $U(1)_{L_e - L_\mu}$ \\ \hline
\end{tabular}\caption{Fate of the symmetries of the Standard Model after gauging a Lepton family difference symmetry.}\label{tab:globMuTau}
\end{table}}

Let us emphasize two essential points:
\begin{itemize}
    \item While the anomaly free combination which remains a standard invertible global symmetry is uniquely fixed, any linearly independent combination generates a non-invertible symmetry and Table \ref{tab:globMuTau} indicates only one possible choice. 
    \item The symmetries enumerated in Table \ref{tab:globMuTau} represent the largest possible symmetry group of this class of models.  A given effective field theory may break some of these symmetries.
\end{itemize}

It is interesting to consider breaking any invertible global symmetries that would constrain the pattern of entries in $y_{ij}^N$, but leaving intact the non-invertible symmetry factor which sets them all identically to zero. In particular the crucial factor in the case of the Standard Model matter content is
\begin{equation}\label{z3s}
    \mathbb{Z}_{N_{g}}^{L}\subset U(1)_{L}~,
\end{equation}
which is a subset of the symmetries in Table \ref{tab:globMuTau} due to the relation \eqref{z3subs}. This is the non-invertible symmetry which will be violated in Section \ref{sec:majModel} by instantons to produce a tiny overall neutrino mass scale; the other global symmetry factors are involved in the `texture' of the Yukawa matrix.

From a top-down perspective, a theory embedding $U(1)_{L_\mu - L_\tau}$ in a group $G$ may spontaneously break the extraneous symmetry factors at the scale where $G$ is broken.
From the bottom up, we may explicitly realize the pattern of symmetry \eqref{z3s} by including in the Lagrangian higher-dimensional operators.  For instance a four-fermion operator of the form: 
\begin{equation}\label{fourfermiop}
    \mathcal{L}\supset \frac{1}{\Lambda^{4}}HL_{e}HL_{e}\bar{e}_{\mu}\bar{e}_{\tau}~,
    \end{equation}
is consistent with $L_{\mu}-L_{\tau}$ gauge invariance and preserves $\mathbb{Z}_{N_g}^L$, but violates the larger possible symmetry of this class of models. Similarly, we may also contemplate six fermion operators of the schematic form:
\begin{equation}
    \mathcal{L}\supset \frac{1}{\Lambda^{11}} (\tilde{H}L_{\mu})^{3}(\tilde{H}L_{\tau})^{3}~,
\end{equation}
which have similar effects on the pattern of symmetry realization in this range of scales. 

The fact that $\mathbb{Z}_{N_{g}}^{L}$ is a non-invertible symmetry can be made more tangible by considering the associated symmetry defect.  Physically this is the topological domain wall which implements the symmetry action on operators.  This action depends dramatically on the kind of operator in question.  On local operators, we simply see a third root of unity Lepton number rotation.  

\begin{figure}[h]
    \centering
    \begin{tikzpicture}
    \draw (0,0)  node[circle,
                       shade,
                       ball color=orange,
                       minimum size=2.3cm,opacity=.5]{};
    \draw[fill,Green] circle(.1);
    \foreach \x in {0, 45, ..., 315}
        {
            \draw[Green,->] (\x:.4) -- ++(\x:.5);
        }
    \node[anchor=south east] at (0,0) {$T$};
    \node[cross out,draw,Blue,minimum size=5,inner sep=0pt,thick] at (60:.9) {};
    \node at (60:1.5) {$\exp\left(i \int C_{0}dt\right)$};
    \node at (-1.2,1.2) {$t=t_0$};
    \end{tikzpicture}
    \caption{A time slice where the non-invertible symmetry defect $\mathcal{D}_{kN}$ (orange) wraps the 't Hooft line, or monopole, $T$ (green). The particle $T$ emits a magnetic flux $\int F_{\mu\nu}\mathrm{d}S^{\mu\nu}$ indicated by the arrows. From the defect Lagrangian \eqref{eq:defect_Lagrangian}, we see that this flux effectively induces a Wilson line $\exp(i \int C_{0}dt)$ of the defect gauge field $C_{\mu}$ stretching along time.  This Wilson line is the worldline of an anyon.}
    \label{fig:my_label}
\end{figure}
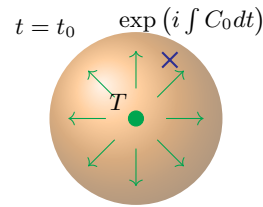

However, when acting on extended objects the behavior of the domain wall is richer.  Indeed, as compared to a standard (invertible) global symmetry, this defect hosts  non-trivial topological field theory, in this case the $3d$ Chern-Simons theory at level $N_g$, $U(1)_{N_{g}}$.  The anyonic degrees of freedom of the defect are then coupled to the bulk $4d$ physics as in the physics of the fractional hall effect: the one-form symmetry on the defect worldvolume is gauged in the bulk.   More concretely, letting $A'$ denote the dynamical gauge field for $L_{\mu}-L_{\tau},$ and $C$ the dynamical Chern-Simons gauge field supported along the defect worldvolume, we have the following defect Lagrangian:
\begin{equation}\label{eq:defect_Lagrangian}
    \mathcal{L}= \frac{iN_{g}}{4\pi}\int_{\Sigma} C_{\mu} \partial_{\nu}C_{\sigma}\varepsilon^{\mu\nu\sigma}d^{3}x +\frac{i}{2\pi}\int_{\Sigma} C_{\mu}\partial_{\nu}A'_{\sigma}\varepsilon^{\mu\nu\sigma}d^{3}x~.
\end{equation}
In particular, this means that an 't Hooft line of the $Z'$ gauge group, physically the worldline of a heavy magnetic monopole of $L_{\mu}-L_{\tau},$ excites the anyons when placed in the $\mathbb{Z}_{3}^{L}$ symmetry defect. See Figure \ref{fig:my_label}.  Such anyons have abelian statistics with fractional spin
\begin{equation}
    \text{spin}=\frac{1}{2N_{g}}~.
\end{equation}
Here the spin $s$ of an anyon is defined by the phase of the wave function obtained after a rotation by angle $2\pi$:
\begin{equation}
    R_{2\pi} |\text{anyon}\rangle = e^{2\pi i s}|\text{anyon}\rangle~.
\end{equation}
The definition is the same as for a familiar boson or fermion, but for anyons the spin can take values other than a half integer. If we move the symmetry defect through the 't Hooft line, it acquires a fractional $L_{\mu}-L_{\tau}$ charge of $1/N_{g}$ by the Witten effect and hence is attached to a topological surface operator. See Figure \ref{fig:D_on_T}.

Finally, let us reconsider the possibility of Majorana neutrino masses in this class of models.  At first pass, a standard effective field theory analysis would make use of only the invertible symmetry to forbid operators from the effective action.  However, from \eqref{z3s} we see that the only invertible symmetry here is contained in $U(1)_{B}$ and hence permits arbitrary Majorana neutrino masses encoded by the Weinberg operator.  

If instead, the non-invertible symmetry $\mathbb{Z}_{N_{g}}^{L}$ is remembered, the situation is qualitatively different: the Weinberg operator is charged under the non-invertible symmetry and excluded from the effective action.  Therefore in this class of models, neutrino masses will be naturally small provided that $\mathbb{Z}_{N_{g}}^{L}$ remains an approximate non-invertible symmetry.

\subsection{Inclusion of Right-Handed Neutrinos}

For a model of Dirac neutrino masses we add three right-handed neutrinos $N_i$ to the Standard Model, and we remind that, as above in Table \ref{tab:charges}, the right-handed neutrinos carry lepton number $L_i$. With these additional light, charged fields in our theory the symmetry analysis of Section \ref{sec:SMsyms} is then incomplete. These neutrinos begin with the global flavor symmetry $U(N_g)_N$, and since they are `sterile' this symmetry is not broken by instanton effects of the Standard Model gauge group. If we add no more explicit global symmetry violation, we have left-handed and right-handed neutrinos which are massless and have nothing to do with each other. 

When we now gauge $L_\mu - L_\tau$ to repeat the analysis of Section \ref{sec:gaugedMuTau} this explicitly breaks the non-Abelian parts of the neutrino flavor symmetry as each now has different quantum numbers. Classically, we should then discuss instead $U(1)_{N_e}\times U(1)_{N_\mu} \times U(1)_{N_\tau}$. These of course have no mixed anomalies with the SM gauge group, but they do have nontrivial ABJ anomalies with $U(1)_{L_\mu - L_\tau}$, as seen in Table \ref{tab:anomMuTauDirac}. 

\begin{table}[h]\centering
\large
\begin{tabular}{|c|c|c|c|c|}  \hline
 & $U(1)_{L_{e} - L_{\mu}}$ & $U(1)_{N_{e}}$  & $U(1)_{N_{\mu}}$&$U(1)_{N_{\tau}}$ \\ \hline

$U(1)_{L_{\mu}-L_{\tau}}^{2}$ & $0$ & $0$ & $1$ & $1$\\ \hline 

\end{tabular}\caption{Mixed anomalies of classical global symmetries of the Standard Model and right-handed neutrinos with a gauged $U(1)_{L_{\mu}-L_{\tau}}$.}\label{tab:anomMuTauDirac}
\end{table}

Clearly $U(1)_{N_e}$ should be anomaly-free, as is $U(1)_{N_\mu - N_\tau}$, as now is $U(1)_{L_e - L_\mu}$ with the inclusion of the right-handed neutrinos. 
However, since we have not introduced any interactions between the right-handed neutrinos and the Standard Model fermions, the global rotations of the previous section---that is, rotations acting on $L_i$ and $\bar e_i$ but not $N_i$---still remain good classical symmetries. But note that these rotations of Table \ref{tab:anomSM2} are no longer proper lepton number symmetries now that we have additional leptons. We will introduce the charge $\tilde{L}_i$ for these symmetries to consistently continue using $L_i$ as a lepton number charge, noting for example that now the electron family number is $L_e = \tilde{L}_e - N_e$. 

Now that we have $L_\mu - L_\tau$ anomalies both for $U(1)_{N_e - N_\mu}$ and for $U(1)_{\tilde{L}_e - \tilde{L}_\mu}$, we find that the linear combination which is the normal lepton family difference symmetry $U(1)_{L_e - L_\mu}$ is now an invertible symmetry. But the orthogonal combination, $\tilde{L_e} - \tilde{L_\mu} + N_e - N_\mu$ has become non-invertible, as laid out in Table \ref{tab:globMuTauDirac}.

{\setlength{\tabcolsep}{0.6 em}
\renewcommand{\arraystretch}{1.3}
\begin{table}[h]\centering
\large
\begin{tabular}{|c|c|}  \hline
Gauged & $U(1)_{L_{\mu}-L_{\tau}}$ \\ \hline

Invertible & $\sfrac{U(1)_{B-N_g N_c \tilde{L}_e}}{\mathbb{Z}_{N_c}}$ \\ 
& $\times U(1)_{N_\mu - N_\tau}$ \\
& $\times U(1)_{L_e - L_\mu}$ \\ 
& $\times U(1)_{N_e}$ \\ \hline

Non-Invertible & $U(1)_{\tilde{L}_e + N_e-\tilde{L}_\mu -N_\mu}$ \\ \hline
\end{tabular}\caption{Fate of the symmetries of the Standard Model $+$ right-handed neutrinos after gauging a lepton family difference symmetry.
}\label{tab:globMuTauDirac}
\end{table}}

As commented above, the symmetries shown in Table \ref{tab:globMuTauDirac} are the largest which may be realized given this spectrum of fermions. To realize a realistic neutrino Yukawa matrix, our UV completion must result in much of this being broken. In particular, in the UV completion of Section \ref{sec:dirModel}, it is only the overall $U(1)_{\tilde{L} + N}$ which is violated solely by instantons, and the rest of the the invertible symmetries may be broken along with the Higgsing $SU(3)_H \rightarrow U(1)_{L_\mu - L_\tau}$.

\subsection{One-Loop Renormalization Group Equations}

We comment on two qualitatively interesting effects arising perturbatively in these gauged $L_\mu - L_\tau$ models.  
The first is the $L_\mu - L_\tau$ beta function, which depends on the existence of the right-handed neutrinos: 
\begin{equation}
\beta(g_{\mu\tau}) = \frac{g_{\mu\tau}^3}{24 \pi^2} \sum_{i=1}^{N_{\rm Weyl}} q_i^2 \Longrightarrow \beta_{\rm M} = \frac{g_{\mu\tau}^3}{4\pi^2}~,\  \beta_{\rm D} = \frac{g_{\mu\tau}^3}{3\pi^2}~,
\end{equation}
where above the subscript M/D indicates the value in the Majorana/Dirac case.

Our neutrino masses will ultimately be generated by a gauge theory effect, the size of which will depend on the size of the gauge coupling at a higher scale. So we emphasize that the discovery of the $L_\mu - L_\tau$ gauge boson and measurement of the `range' of this force $M_{Z'}$ along with its strength $g^2_{\mu\tau}(M_{Z'}^2) \equiv g_0^2$ allows us to evolve the gauge theory up to higher scales, defining as usual $\alpha_{\mu\tau}(\mu^2) \equiv \frac{g^2_{\mu\tau}(\mu^2)}{4 \pi}$,
\begin{equation} \label{eqn:mutauRunning}
    \alpha_{\mu\tau}(\mu^2)^{-1} = \alpha_{0}^{-1} - \frac{c_i}{\pi} \log \frac{\mu^2}{M_{Z'}^2}~,
\end{equation}
with $c_{\rm M} = 1$ and $c_{\rm D} = 4/3$,
revealing the presence of a Landau pole for $\mu^2 \sim M^2_{Z'} \exp \frac{4 \pi^2}{c_i g^2_0}$. For $g_0 \gtrsim 1$, as will be relevant for us below, the Landau pole is far closer than the familiar one in the hypercharge coupling, so embedding $L_\mu - L_\tau$ in a non-Abelian gauge group is quite well-motivated on general grounds.

Finally, there is one more marginal gauge-invariant operator in this theory, which is kinetic mixing $\epsilon B^{\mu\nu} F'_{\mu\nu}$. The presence of matter charged under both $U(1)_Y$ and $U(1)_{L_\mu - L_\tau}$ generically generates such kinetic mixing at one loop.  
However, since the second and third generations have the same hypercharges and opposite $L_\mu - L_\tau$, the divergent parts of the one-loop diagrams cancel.
Then without any new light charged fields, kinetic mixing will not be induced until after electroweak symmetry-breaking and below the mass of the tau lepton, where there will be mixing of the $Z'$ with the photon of size
\begin{equation}
    \epsilon \sim \frac{e g_{\mu\tau}}{32 \pi^2} \ln \frac{m_\tau^2}{m_\mu^2}~.
\end{equation}
So we need not worry about kinetic mixing if we stick to massive $Z'$s above the scales which have been probed already at colliders, but in a broader investigation including lighter $L_\mu - L_\tau$ gauge bosons, for example to address $(g-2)_\mu$, this would impact the phenomenology (see \cite{Hook:2010tw}).

\section{Majorana Mass Model} \label{sec:majModel}

Having shown the Standard Model extended by $U(1)_{L_\mu - L_\tau}$ has an exact non-invertible global symmetry which forbids neutrino masses, we now set about evincing a UV completion to break this symmetry and provide the observed small neutrino masses. 
Superficially, thus far our work echos much of the broad literature on models in which a symmetry protects neutrino masses. However, we need not introduce any separate spontaneous symmetry-breaking sector nor any explicit breaking communicated via mediators with prescribed charges under some new global symmetries.

\begin{table}[h]\centering
\large
\begin{tabular}{|c|c|c|c|c|c|c|}  \hline
 & $SU(2)_{H}$ & $U(1)_Z$ & $L_\mu-L_\tau$ & $U(1)_L$ \\ \hline

$\mathbf{\Phi}$ & $2$ & $-1$ & $\begin{pmatrix} \Phi_e \\ \Phi_\tau \end{pmatrix} = \begin{pmatrix} 0 \\ -1 \end{pmatrix}$ & $0$ \\ \hline

$\mathbf{L}_{\mu e}$ & $2$ & $+1$ & $\begin{pmatrix} L_\mu \\ L_{e_1} \end{pmatrix} = \begin{pmatrix} +1 \\ 0 \end{pmatrix}$ & $+1$ \\ \hline
$\mathbf{L}_{E \tau}$ & $2$ & $-1$ & $\begin{pmatrix} L_{e_2} \\ L_\tau \end{pmatrix} = \begin{pmatrix} 0 \\ -1 \end{pmatrix}$ & $+1$ \\ \hline
$\psi_L$ & -- & $0$ & $0$ & $-1$ \\ \hline

$\mathbf{\bar e}_{\mu e}$ & $2$ & $-1$ & $\begin{pmatrix} \bar{e}_1 \\ \bar \mu \end{pmatrix} = \begin{pmatrix} 0 \\ -1 \end{pmatrix}$ & $-1$ \\ \hline
$\mathbf{\bar e}_{E \tau}$ & $2$ & $+1$ & $\begin{pmatrix} \bar \tau \\ \bar{e}_2 \end{pmatrix} = \begin{pmatrix} +1 \\ 0 \end{pmatrix}$ & $-1$ \\ \hline
$\psi_{\bar e}$ & -- & $0$ & $0$ & $+1$ \\ \hline

\end{tabular}\caption{Fields and their representations under the relevant symmetry groups. $\psi_L$ and $\psi_{\bar e}$ are in conjugate representations of the SM gauge group with respect to $L$ and $\bar e$ so that the extra fields are overall vector-like with respect to $G_{SM}$.}\label{tab:chargesMaj}
\end{table}

Instead, the fact that it is a non-invertible symmetry protecting neutrino masses here means that this symmetry can be broken in the gauge sector itself by purely quantum effects, and so naturally be broken an exponentially small amount. To provide such a Majorana mass we consider a UV model in which the lepton family difference is embedded into a horizontal symmetry $SU(2)_H \times U(1)_Z$ \footnote{
We note that this model does not admit finite mass monopole solutions. This is related to the fact that the UV model has a magnetic one-form symmetry whose current flows to the IR magnetic one-form symmetry current.
Nevertheless, we only need virtual states to run in the loop in Figure \ref{fig:mon_loop}. 
Here, the relevant unstable virtual state is the one having the same asymptotic gauge field profile as the on-shell 't Hooft-Polykov monopole in the related model of adjoint Higgsing $SU(2) \to U(1)$, and the coupling of the neutrino to such a state is determined by the UV anomaly coefficient.
In the case of breaking $SU(2)\times U(1)\to U(1)$, this gauge field profile is null-homotopic and thus unstable, but the estimate of the magnitude of contribution made in \eqref{eq:mon_inst} applies. If we further embed the $SU(2)\times U(1)$ gauge group into $SU(3)$, the unstable state can be understood as a bound state a monopole and an anti-monopole with different moduli.}. At the group level this mirrors the emersion of QED out of the electroweak sector, and to our knowledge has not been studied before. In particular, it is different from the oft-studied $SU(2)_H$ UV completion with lepton species in adjoint irreducible representations \cite{He:1991qd} and instead embeds each lepton species into the reducible $2 \oplus 2$ of $SU(2)_H$. We may easily lift the one extra generation of each species by including a conjugate $SU(2)_H$ singlet such that the total addition of matter is vector-like under the SM gauge groups, resulting in a theory that bears some resemblance to $SU(2)_H$ theories putting the leptons in a $2 \oplus 1$ where the electron is a singlet \cite{Chiang:2017vcl,Heeck:2011wj}.

In this case our UV theory contains the fields of Table \ref{tab:chargesMaj} which interact as follows
\begin{align}\label{eq:Maj_Yuk}
    \mathcal{L} &\supset y_\mu H \mathbf{L}_{\mu e} \mathbf{\bar e}_{\mu e} + y_\tau H \mathbf{L}_{E \tau} \mathbf{\bar e}_{E \tau} 
    + \lambda_{L_1} \bm{\Phi} \mathbf{L}_{\mu e} \psi_L\\ & + \lambda_{L_2} \bm{\tilde{\Phi}} \mathbf{L}_{E \tau} \psi_L 
    + \lambda_{e_1} \bm{\tilde{\Phi}} \mathbf{\bar e}_{\mu e} \psi_{\bar e} + \lambda_{e_2} \bm{\Phi} \mathbf{\bar e}_{E \tau} \nonumber\psi_{\bar e}~.
\end{align}
That $U(1)_L$ is the only global symmetry can be checked by using $U(1)_Y$ and $U(1)_Z$ to set the $H$ and ${\bf{\Phi}}$ charges to zero, and then solving the six constraints from the Yukawa operators. In the $SU(2)_H\times U(1)_Z$ breaking when $\bm{\Phi}$ gets a vev $\langle \bm{\Phi} \rangle = (v_\Phi, 0)^\intercal$ , one linear combination of each of the left-handed and right-handed electrons will be lifted by the $SU(2)_H$ singlets. Defining two rotation angles $\tan \theta_L = -\lambda_{L_1}/\lambda_{L_2}$ and $\tan \theta_e = -\lambda_{e_1}/\lambda_{e_2}$, the rotation to the mass basis is
\begin{align}
    \begin{pmatrix} L_e \\ L_E \end{pmatrix}
    &= \begin{pmatrix} \cos \theta_L & -\sin \theta_L \\ \sin \theta_L & \cos \theta_L \end{pmatrix}
    \begin{pmatrix}
    L_{e_1} \\ L_{e_2}
    \end{pmatrix}~, \label{eq:Lrotate} \\
    \begin{pmatrix} \bar e \\ \bar E \end{pmatrix}
    &= \begin{pmatrix} \cos \theta_e & -\sin \theta_e \\ \sin \theta_e & \cos \theta_e \end{pmatrix}
    \begin{pmatrix}
    \bar{e}_1 \\ \bar{e}_2
    \end{pmatrix}~,
    \label{eq:erotate}
\end{align}
where $L_E$ and $\bar E$ pair up with $\psi_L, \psi_{\bar e}$ to get large Dirac masses $M_L = \sqrt{\lambda_{L_1}^2 + \lambda_{L_2}^2} v_\Phi$ and $M_e = \sqrt{\lambda_{e_1}^2 + \lambda_{e_2}^2} v_\Phi$.
The other linear combination remains as the light electron and in the infrared, at lowest order (in $v/v_\Phi$) we have
\begin{align}
    \mathcal{L} &\supset y_\mu H L_\mu \bar \mu + y_\tau H L_\tau \bar \tau \nonumber \\ 
    &+\left(y_\tau \sin \theta_L \sin \theta_e + y_\mu \cos \theta_L \cos \theta_e \right) H L_{e} \bar e~,
\end{align}
with more than enough freedom to match the electron mass, though with some fine-tuning required. 
We neglect to write down the additional Yukawa couplings of the Higgs with the extra vector-like fermions, but mention that the subleading, off-diagonal Yukawas are also constrained through SMEFT by low-energy and precision data \cite{Ellis:2020unq,Ethier:2021bye}.

\begin{table}[h]\centering
\large
\begin{tabular}{|c|c|c|c|}  \hline
 & $SU(2)_L^2$ & $U(1)_Y^2$ & $U(1)_{L_\mu-L_\tau}^2$ \\ \hline

$U(1)_L$ & $N_g$ & $-18 N_g$ & $(N_g - 1)$ \\ \hline \hline

 & $SU(2)_{H}^2$ & $U(1)_Z^2$ & \\ \hline

$U(1)_L$ & $(N_g - 1)$ & $2(N_g - 1)$ &  \\ \hline

\end{tabular}\caption{Mixed anomalies of $U(1)_L$ in the Majorana mass scenario in the IR (above) and UV (below). }\label{tab:anomMaj}
\end{table}

\begin{figure}[t]
    \centering
        {\includegraphics[clip, trim=0.0cm 0.0cm 0.0cm 0.0cm, width=0.5\textwidth]{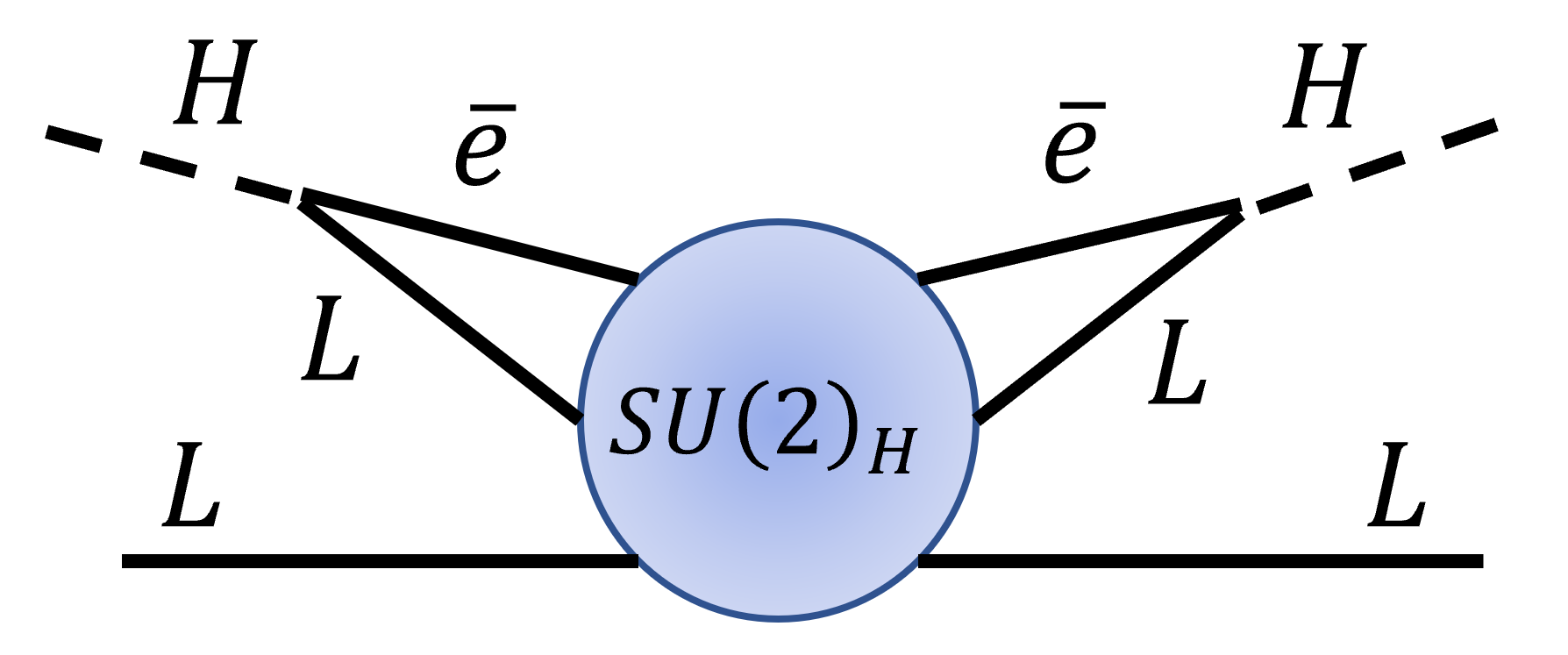}}
    \caption{The 't Hooft vertex generated by $SU(2)_H$ instantons. Two pairs of fermion legs are contracted with the charged lepton Yukawa coupling to the Higgs, which generates a Majorana neutrino mass.}
    \label{fig:MajInst}
\end{figure}

The $U(1)_L \times SU(2)_H^2$ anomaly coefficient in Table \ref{tab:anomMaj} shows $U(1)_L$ is broken down to $\mathbb{Z}_2^L \subset U(1)_L$ by $SU(2)_H$ instantons, which allows Majorana mass terms. Recalling that the exact non-invertible symmetry at intermediate scales is the subset $\mathbb{Z}_{N_g}^L \subset U(1)_{L}$ protecting neutrino masses in the $U(1)_{L_\mu - L_\tau}$ gauge theory, we see that in the UV theory $U(1)_L$ is then entirely broken by quantum effects. Operationally, the breaking by $SU(2)_H$ suffices to produce the effect of interest, since $SU(2)_H^2$ instantons generate 't Hooft vertices as

\begin{equation}
\mathcal{L} \sim \frac{e^{-\frac{2\pi}{\alpha_H}}}{v_\Phi^5} L_{\mu e} L_{\mu e} L_{E \tau} L_{E \tau} \bar{e}_{\mu e} \bar{e}_{E\tau}~,
    \label{eq:maj_vertex}
\end{equation}
where the exponential suppression is the action of the instanton as discussed above. From this operator, we may immediately use the charged lepton Yukawa couplings to contract two pairs of charged fermion legs into Higgses as in Figure \ref{fig:MajInst} and have
\begin{equation}
\mathcal{L} \sim \frac{y_\mu y_\tau}{v_\Phi} e^{-\frac{2\pi}{\alpha_H}} \tilde{H} L_{\mu e} \tilde{H} L_{E \tau}~.
\end{equation}
While the IR-dominance of the instantons means that the largest contributions come from the energy at which $SU(2)_H \times U(1)_Z$ is broken, the instantons themselves are an effect of $SU(2)_H$ gauge theory and so respect this UV gauge symmetry. 

To lowest order, the neutrino masses generated at $v_\Phi$ only see their breaking through the lifting of the vector-like partners, as
\begin{align}
    \mathcal{L} &\sim \frac{y_\mu y_\tau}{v_\Phi} e^{-\frac{2\pi}{\alpha_H}} \left[ \tilde{H} L_\mu \tilde{H} L_\tau -  \tilde{H} L_{e_1} \tilde{H} L_{e_2} \right] \nonumber \\ &+  v_{\Phi} \left( \lambda_{L_1} L_{e_1} - \lambda_{L_2} L_{e_2}\right) \psi_L ~,
\end{align}
where the second line contains Dirac masses both for the charged lepton above and for one neutrino. 
Again to lowest order this just removes the corresponding $\nu_E$ from the spectrum and we have
\begin{equation} \label{eqn:majMass}
    \mathcal{L} \sim y_\mu y_\tau \frac{v^2}{v_\Phi} e^{-\frac{2\pi}{\alpha_H}} \Bigl[ \nu_\mu \nu_\tau - \half \sin 2 \theta_L \nu_{e} \nu_{e} \Bigr]~.
\end{equation}
We see explicitly now that our UV completion of the $U(1)_{L_\mu - L_\tau}$ theory does not respect the $U(1)_{B-N_g N_c L_e}$ global symmetry we observed above that the $U(1)_{L_\mu - L_\tau}$ extension of the Standard Model could enjoy.  
In particular, Higgsing of our ultraviolet gauge theory generates four fermion operators such as \eqref{fourfermiop}, while in the neutrino mass matrix it generates $L_{e}$ violating entries in conjunction with the 't Hooft vertex \eqref{eq:maj_vertex}. This is desirable, as while the SM proper respects this symmetry the observed neutrino phenomenology does not.

When the $\psi_L$ couplings are near-universal $\lambda_{L_1} \simeq \lambda_{L_2}$ we have $\sin(2 \theta_L) \simeq 1$ and are well set up for a quasi-degenerate neutrino spectrum with large $2-3$ mixing. This successful prediction of the large mixing angle for `atmospheric' neutrinos has long made $L_\mu - L_\tau$ symmetries particularly interesting for neutrino theorists \cite{Foot:1990mn,He:1990pn,He:1991qd,Foot:1994vd}.

So far we have only taken into account the physics above $v_\Phi$, which provides the violation of the non-invertible $Z_{N_g}^L$ symmetry by $SU(2)_H$ instantons. The Higgsing leaves unbroken a remnant $U(1)_{L_\mu - L_\tau}$ subgroup, which prevents us from populating the other entries of the Majorana mass matrix. Since this is manifestly not gauged in the far IR, we must have further spontaneous breaking with a scalar $\phi$ effecting
\begin{equation}
    \left\langle \phi \right\rangle \neq 0 \ \Rightarrow \ U(1)_{L_\mu - L_\tau} \rightarrow \varnothing~.
\end{equation}
This is the same scalar providing mass for the massive $Z'$ boson whose discovery would be the harbinger of this mechanism. Given the mass $M_{Z'}^2$ and the coupling $g_{\mu\tau}(M_{Z'}^2)$, we can express the neutrino mass scale $m_\nu$ using \eqref{eqn:majMass} and the beta function, 
\begin{equation} \label{eqn:nuMassMaj}
    m_\nu \sim \frac{m_\mu m_\tau}{v_\Phi} \left( \frac{v_\Phi}{M_{Z'}}\right)^{4 s_H^2} \exp \frac{-2 \pi s_H^2}{\alpha_{\mu\tau}(M_{Z'}^2)}~,
\end{equation}
where we have plugged in $y_\mu v = m_\mu, y_\tau v = m_\tau$, and we have also introduced the `horizontal mixing angle' $s_H^2 \equiv \sin^2 \theta_H$, where $g_{\mu\tau}(v_\Phi) \equiv g_H(v_\Phi) \sin \theta_H$ in analogy to the electroweak sector. In terms of the microscopic gauge couplings, $s_H = 2 g_Z / \sqrt{g_H^2 + 4 g_Z^2}$, where the factor of two difference from the SM case is due to ${\bf \Phi}$ having unit charge. Then we can easily invert \eqref{eqn:nuMassMaj} to find the needed UV scale as
\begin{equation}\label{eqn:scaleMaj}
    v_{\Phi}^{4 s_H^2-1} \sim M_{Z'}^{4 s_H^2-1} \left(\frac{M_{Z'} m_\nu}{m_\mu m_\tau} \right) \exp \frac{2\pi s_H^2}{\alpha_{\mu\tau}(M_{Z'}^2)}~.
\end{equation}
The $s_H$ dependence reflects that the contribution of the $SU(2)_H$ instantons to violating the non-invertible symmetry depends upon how $U(1)_{L_\mu - L_\tau}$ emerges from $SU(2)_H \times U(1)_Z$ upon Higgsing.

\begin{figure}[h]
    \centering
        {\includegraphics[clip, trim=0.0cm 0.0cm 0.0cm 0.0cm, width=0.5\textwidth]{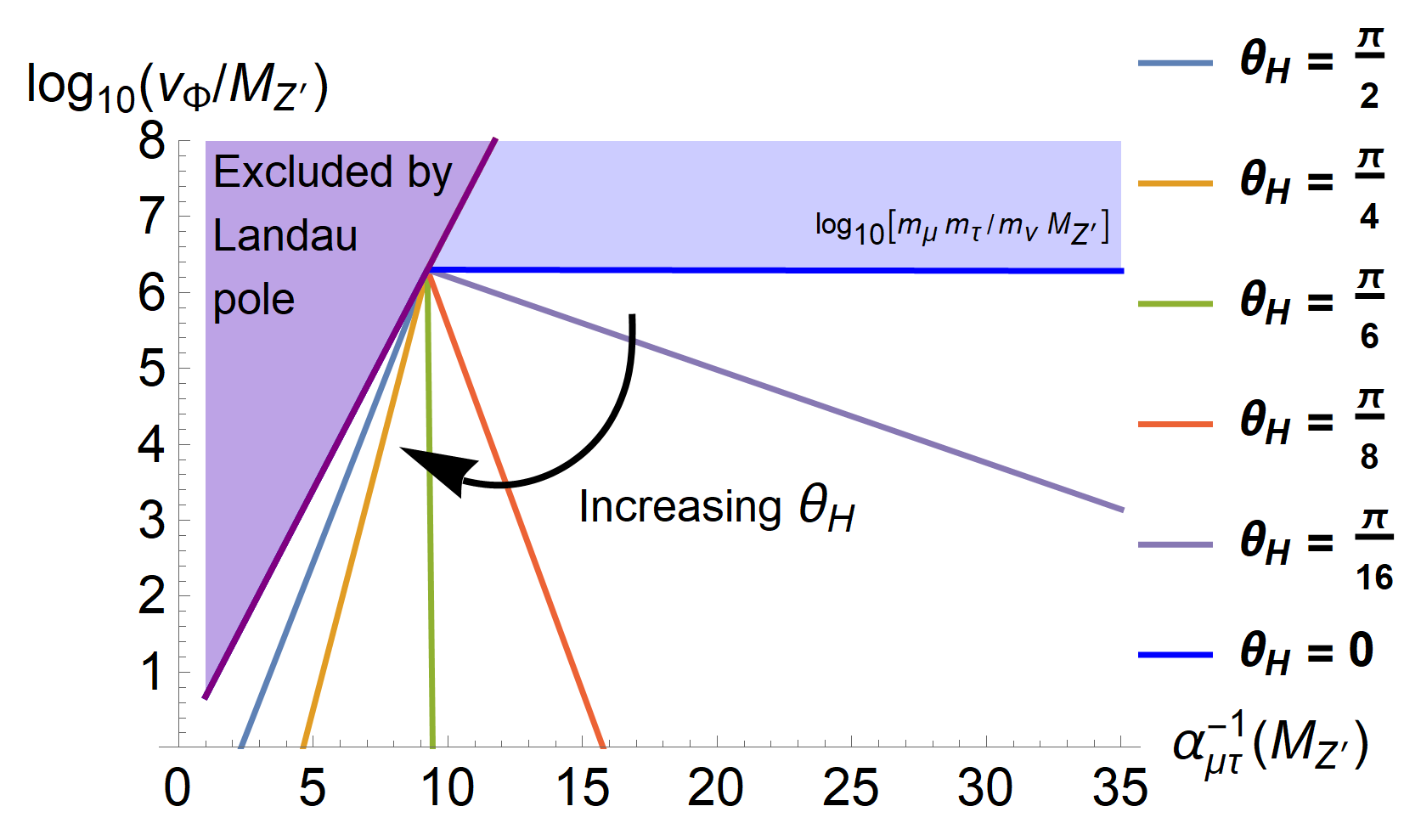}}
    \caption{
    The UV scale $v_\Phi$ predicted upon discovery of a $U(1)_{L_\mu - L_\tau}$ $Z'$ boson at $M_{Z'}=1 \ {\rm TeV}$ over a range of coupling strengths $\alpha_{\mu\tau}^{-1}(M_{Z'}) \equiv 4 \pi / g_{\mu\tau}^2(M_{Z'})$.
    A different value of $M_{Z'}$ merely shifts the plot vertically. The purple line denotes the scale at which the Landau pole in $U(1)_{L_\mu - L_\tau}$ is reached.}
    \label{fig:scaleCompare}
\end{figure}

In contrast to the Dirac case below, here the scale of the UV completion is not uniquely predicted by the low-energy data but is instead partially degenerate with the mixing angle. In Figure \ref{fig:scaleCompare} we plot the UV scale as a function of the inverse coupling strength over the full range of $0 < \sin \theta_H < 1$, which displays several interesting features. It is useful to massage \eqref{eqn:scaleMaj} into the form
\begin{equation}
    \frac{4 s_H^2-1}{s_H^2} \log \frac{v_\Phi m_\nu}{m_\mu m_\tau}
    \sim 4 \log\frac{m_\mu m_\tau}{m_\nu M_{Z'}} + \frac{2\pi}{\alpha_{\mu\tau}(M_{Z'}^2)}~.
\end{equation}
Here one may see clearly that for a given $s_H$ there will be a line with slope $2 \pi s_H^2/(4 s_H^2 - 1)$, and the value of $M_{Z'}$ just translates the line vertically. The limiting value $s_H^2 \rightarrow 0$ is reached at a fixed $\alpha_{\mu\tau}^{-1}(M_{Z'})$ by sending $\alpha_H \rightarrow \infty$. The disappearance of the exponential suppression means this gives uniformly $\log v_\Phi \sim \log m_\mu m_\tau / m_\nu$, but of course our picture of 't Hooft vertices has properly broken down.

At the special value $\theta_H = \pi/6$ ($s_H = 1/2$), the UV scale drops out of \eqref{eqn:nuMassMaj} and this model predicts uniquely a coupling $2 \pi \alpha_{\mu\tau}^{-1} \sim 4 \log M_{Z'} m_\nu / (m_\mu m_\tau)$.
Plugging the equation for the Landau pole into \eqref{eqn:scaleMaj} one finds that the $s_H$ dependence drops out of the relation to leave also this same coupling, so all lines of fixed $s_H$ meet the Landau pole at this same point.

Intriguingly, this is precisely the mixing angle demanded by the unification $SU(2)\times U(1) \subset SU(3)$ \cite{Weinberg:1971nd,Georgi:1974sy}, meaning that a $U(1)_{L_\mu - L_\tau}$ gauge boson satisfying this relationship between $M_{Z'}$ and $\alpha_{\mu\tau}^{-1}$ would be a smoking gun for this model.

Renormalization group flow from $v_\Phi$ to a lower scale generically generates higher-dimensional operators involving $\phi$ which respect the infrared symmetries. Since the theory below $v_\Phi$ contains the non-invertible $\mathbb{Z}_3^L$ symmetry, any further generation of Majorana masses must come from an insertion of the $\mathbb{Z}_3^L$-violating operators already present at the scale $v_\Phi$. For $\phi$ with charge $1$ under the gauged $L_\mu - L_\tau$ but uncharged under the global total lepton number, our infrared Lagrangian should then include 
\begin{align}
    \mathcal{L} \supset y_\mu y_\tau \frac{v^2}{v_\Phi} e^{-\frac{2\pi}{\alpha_H}} \bigl[ & \nu_e \nu_\tau \frac{\phi}{v_\Phi} + \nu_e \nu_\mu \frac{\phi^\dagger}{v_\Phi}  \\&+ \nu_\tau \frac{\phi}{v_\Phi} \nu_\tau \frac{\phi}{v_\Phi} + \nu_\mu \frac{\phi^\dagger}{v_\Phi} \nu_\mu \frac{\phi^\dagger}{v_\Phi}\bigr]~, \nonumber
\end{align}
where we have left off additional $\mathcal{O}(1)$ factors in front of these operators. We give this merely as a schematic for how $L_\mu - L_\tau$ symmetry-breaking effects are generated; fully realistic models will require additional sources of symmetry-breaking for the invertible symmetry factors which also control the Yukawa texture.

There is a large body of work on how to achieve realistic neutrino masses and mixing in the context of gauged $L_\mu - L_\tau$ symmetry, possibly with a stage of $\mu-\tau$ reflection symmetry. As our point here is to evince the generation of the neutrino mass scale from nonperturbative, inherently quantum-mechanical violation of non-invertible $\mathbb{Z}_3^L$, we leave the detailed study of this further soft-breaking to future work, and for now refer to \cite{Binetruy:1996cs,Araki:2019rmw,Joshipura:2019qxz,Ota:2006xr,Mohapatra:2005yu,Fuki:2006xw,Chen:2017gvf,Heeck:2011wj,Ma:2001md,Bell:2000vh,Chamoun:2019pbh,Asai:2018ocx,Asai:2017ryy,Lashin:2013xha,Gupta:2013it,Chun:2007vh,Baek:2015mna,Nomura:2018cle,Dev:2017fdz,Heeck:2022znj,Choubey:2004hn}, among many others, for discussions in this direction. 

\section{Dirac Mass Model} \label{sec:dirModel}

To break the non-invertible symmetry in the Dirac case, we embed $U(1)_{L_\mu - L_\tau}$ into a non-Abelian horizontal $SU(3)_H$ under which the leptons are fundamentals. 
No extra fermions are needed, and this completion of lepton family difference symmetries has recently been studied in \cite{Alonso-Alvarez:2021ktn}. 
We emphasize in particular that we consider gauging only the horizontal lepton flavor as opposed to the horizontal symmetry of simple GUT models which includes both quarks and leptons \cite{Berezhiani:1985in}.

\begin{table}[h]\centering
\large
\begin{tabular}{|c|c|c|c|c|}  \hline
 & $SU(3)_{H}$ & $U(1)_{\mu-\tau}$ & $U(1)_{\tilde{L}}$ & $U(1)_N$\\ \hline

$\mathbf{L}$ & $3$ & $\begin{pmatrix} L_e \\ L_\mu \\ L_\tau \end{pmatrix} = \begin{pmatrix} 0\\ +1 \\ -1 \end{pmatrix}$ & $+1$ & $0$\\ \hline

$\mathbf{\bar e}$ & $\bar 3$ & $\begin{pmatrix} \bar e \\ \bar \mu \\ \bar \tau \end{pmatrix} = \begin{pmatrix} 0 \\ -1 \\ +1 \end{pmatrix}$ & $-1$ & $0$\\ \hline

$\mathbf{N}$ & $\bar 3$ & $\begin{pmatrix} N_e \\ N_\mu \\ N_\tau \end{pmatrix} = \begin{pmatrix} 0 \\ -1 \\ +1 \end{pmatrix}$ & $0$ & $+1$ \\ \hline

\end{tabular}\caption{Fields and their representations under the relevant symmetry groups. }\label{tab:chargesDirac2}
\end{table}

The matter content in the lepton sector is given in Table \ref{tab:chargesDirac2}, and the Lagrangian of the UV theory is simply
\begin{equation}
    \mathcal{L} = y_\tau H \mathbf{L} \mathbf{\bar e}~.
\end{equation}
The matter fields at the level of the gauge theory enjoy the global symmetry $U(1)_{\bf{L}} \times U(1)_{\bf{\bar e}} \times U(1)_{\bf N}$, and the charged lepton Yukawa coupling breaks one combination explicitly $U(1)_{\bf{L}} \times U(1)_{\bf{\bar e}} \rightarrow U(1)_{\tilde{L}}$ as in the theory at intermediate scales. 

\begin{figure}[h]
    \centering
        {\includegraphics[clip, trim=0.0cm 0.0cm 0.0cm 0.0cm, width=0.5\textwidth]{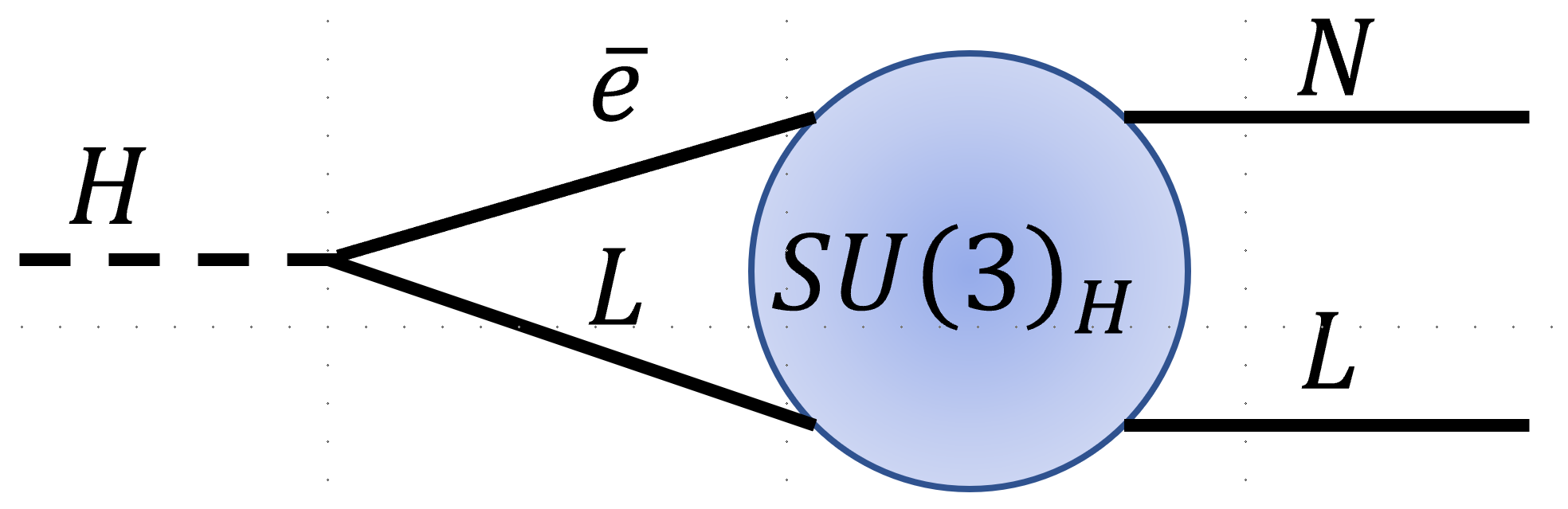}}
    \caption{The 't Hooft vertex generated by $SU(3)_H$ instantons. Two fermion legs are contracted with the charged lepton Yukawa coupling, yielding a Dirac neutrino mass.}
    \label{fig:DiracInst}
\end{figure}

\begin{table}[h]\centering
\large
\begin{tabular}{|c|c|c|c|}  \hline
 & $SU(2)_L^2$ & $U(1)_Y^2$ & $U(1)_{L_\mu-L_\tau}^2$  \\ \hline

$U(1)_{\tilde{L}_e - \tilde{L}_\mu}$ & $0$ & $0$ & $-1$ \\ \hline
$U(1)_{N_e - N_\mu}$ & $0$ & $0$ & $-1$ \\
\hline \hline

 & $SU(3)_{H}^2$ &  & \\ \hline
 
$U(1)_{\tilde{L}}$ & $+1$ &  & \\ \hline 

$U(1)_{N}$ & $+1$ &  & \\ \hline 

\end{tabular}\caption{Mixed anomalies of flavor symmetries in the Dirac mass scenario in the IR (above) and UV (below). }\label{tab:anomDirac3}
\end{table}

As a theory of just three species of gauge fundamentals, it is easy to compute the anomalies in Table \ref{tab:anomDirac3}. 
One finds that the anomaly-free linear combination of the remaining global symmetries is $U(1)_L$, i.e.\ the normal lepton number symmetry, which prevents the production of Majorana masses for the right-handed neutrinos.
The other direction, $\tilde{L}+N$, is anomalous and the associated 't Hooft vertex of Figure \ref{fig:DiracInst} clearly violates both $U(1)_{\tilde{L}}$ and $U(1)_N$ by one unit each, 
\begin{equation}
    \mathcal{L} \sim \frac{e^{-\frac{2\pi}{\alpha_H}}}{v_\Phi^{2}}\mathbf{L} \mathbf{\bar e} \mathbf{L} \mathbf{N}~.
\end{equation}
A single insertion of the charged lepton Yukawa turns this into a Dirac neutrino mass of a size
\begin{equation}\label{eqn:dirMass}
    \mathcal{L} \sim y_\tau e^{-\frac{2\pi}{\alpha_H}} \tilde{H} \mathbf{L} \mathbf{N}~.
\end{equation}
We must still address the generation of the neutrino texture, but we already have the information we need to link low-energy observables to the scale of $SU(3)_H$-breaking. The matching of gauge couplings at $v_\Phi$ is now $4 g_{\mu\tau}^2 = g_H^2$, so we may write the neutrino mass scale as
\begin{equation}
    m_\nu \sim m_\tau \left(\frac{v_\Phi}{M_{Z'}} \right)^{\sfrac{4}{3}} \exp \frac{- \pi}{2 \alpha_{\mu\tau}(M_{Z'}^2)}~.
\end{equation}
Then we can again invert this to find the UV scale $v_\Phi$ given the neutrino mass scale and the measurements of $M_{Z'}$ and $\alpha_{\mu\tau}(M_{Z'}^2)$, 
\begin{equation}\label{eqn:scaleDirac}
    v_{\Phi}^2 \sim M_{Z'}^2 \left(\frac{m_\nu}{m_\tau}\right)^{\sfrac{3}{2}} \exp{\frac{3 \pi}{4  \alpha_{\mu\tau}(M^2_{Z'})}}~, 
\end{equation}
where we recall $\sfrac{m_\nu}{m_\tau}\sim 10^{-11}$. 
By requiring that $M_{Z'} \lesssim v_{\Phi} \lesssim M_{\rm pl}$ we get two-sided limits on the coupling strengths for which this mechanism may work:
\begin{equation}
    \frac{3\pi}{4} \left[\ln \left(\frac{M_{\rm pl}^2}{M_{Z'}^2} \frac{m_\tau^{3/2}}{m_\nu^{3/2}}\right) \right]^{-1}  <  \alpha_{\mu\tau}(M_{Z'}^2) < \frac{\pi}{2} \left[ \ln \frac{m_\tau}{m_\nu} \right]^{-1}~.
\end{equation}
The lower limit depends on the scale at which the $Z'$ is discovered, but for around the TeV scale this limits roughly $\sfrac{1}{45} \lesssim \alpha_{\mu\tau}(M_{Z'}^2) \lesssim \sfrac{1}{16}$. The Dirac theory is then far more predictive by virtue of involving no unknown mixing angles.

Now having discovered a hierarchically small neutrino mass scale from the charged leptons and gauge theory dynamics, at lower energies we must generate the observed neutrino texture. This should originate from the scalar sector responsible for spontaneous symmetry-breaking $SU(3)_H\rightarrow U(1)_{L_\mu - L_\tau}$ and later $U(1)_{L_\mu - L_\tau}\rightarrow \varnothing$, similarly to the recent analysis of \cite{Alonso-Alvarez:2021ktn}.

As we now have a multitude of global symmetries shaping the structure of the Yukawa matrix, there are many more options for combinations of charges to assign to spurions. For a rough analysis of the expected sizes of Yukawa entries, we shall just display which charges are violated by each operator in Table \ref{tab:YukDirac}.

{\renewcommand{\arraystretch}{1.5}
\begin{table}[h]\centering
\small
\begin{tabular}{|c|c|c|c|}  \hline
& {\large $N_e$} & {\large $N_\mu$} & {\large $N_\tau$} \\ \hline

& $\tilde{L}_e = +1$ & ${\mathbf{L_\mu - L_\tau}} = -1$ & ${\mathbf{L_\mu - L_\tau}} = +1$ \\
{\large $L_e$} & $N_e = +1$ & $\tilde{L}_e = +1$ & $\tilde{L}_e = +1$ \\
& & $N_\mu - N_\tau = +1$ & $N_\mu - N_\tau = -1$ \\ \hline

& ${\mathbf{L_\mu - L_\tau}} = +1$ &  & ${\mathbf{L_\mu - L_\tau}} = +2$ \\
{\large $L_\mu$} & $N_e = +1$ & $N_\mu - N_\tau = +1$ & $N_\mu - N_\tau = -1$ \\
& $L_e - L_\mu=-2$ & & $L_e - L_\mu=-1$ \\ \hline

& ${\mathbf{L_\mu - L_\tau}} = -1$ & ${\mathbf{L_\mu - L_\tau}} = -2$ &  \\
{\large $L_\tau$} & $N_e = +1$ & $N_\mu - N_\tau = +1$ & $N_\mu - N_\tau = -1$ \\
& $L_e - L_\mu=-1$ & $L_e - L_\mu=+1$ &  \\ \hline

\end{tabular}\caption{Symmetries protecting various Yukawa entries in the basis used in Table \ref{tab:globMuTauDirac}. 
The non-invertible symmetry protects all entries and its breaking requires a 't Hooft vertex insertion.
The bolding reminds that $L_\mu - L_\tau$ is gauged, so its breaking is on a different footing than the other charges.}\label{tab:YukDirac}
\end{table}}



\let\oldaddcontentsline\addcontentsline
\renewcommand{\addcontentsline}[3]{}
\section*{Acknowledgements}
\let\addcontentsline\oldaddcontentsline

We thank Jeff Harvey and Liantao Wang for useful discussions related to this work.
CC is supported by the US Department of Energy DE-SC0009924 and the Simons Collaboration on Global Categorical Symmetries.
The work of SH at the University of Chicago is supported by the DOE grant DE-SC-0013642. SH is partially supported by the U.S. Department of Energy under contracts No.~DE-AC02-06CH11357 at  Argonne National Laboratory. SH is supported by Basic Research Support Grant at the Korea Advanced Institute of Science and Technology (KAIST).
SK is supported by an Oehme Postdoctoral Fellowship from the Enrico Fermi Institute at the University of Chicago. 
KO is supported by JSPS KAKENHI Grant-in-Aid No.22K13969 and the Simons Collaboration on Global Categorical Symmetries.
SH and SK are grateful for the hospitality of the Aspen Center for Physics, where this work was performed in part, which is supported by National Science Foundation grant PHY-1607611.


\let\oldaddcontentsline\addcontentsline
\renewcommand{\addcontentsline}[3]{}
\bibliography{noninv}
\let\addcontentsline\oldaddcontentsline

\end{document}